\documentclass[accepted]{uai2023} 

\usepackage[american]{babel}

\usepackage{natbib}
\usepackage{mathtools}
\usepackage{booktabs}
\usepackage{tikz}

\usepackage{algorithm2e}
\usepackage{amsmath}
\usepackage{amssymb}
\usepackage{mathtools}
\usepackage{amsthm}
\usepackage{bbm}
\usepackage{bm}
\usepackage{array}
\usepackage[graphicx]{realboxes}
\usepackage{multirow}

\newcommand{\x}{\mathbf{x}}
\newcommand{\X}{\mathbf{X}}

\DeclareMathOperator\supp{supp}

\theoremstyle{plain}
\newtheorem{theorem}{Theorem}[section]

\theoremstyle{definition}
\newtheorem{definition}[theorem]{Definition}

\theoremstyle{remark}

\newtheorem{manualtheoreminner}{Theorem}
\newenvironment{manualtheorem}[1]{%
  \renewcommand\themanualtheoreminner{#1}%
  \manualtheoreminner
}{\endmanualtheoreminner}

\title{Variable Importance Matching for Causal Inference}
\author[1]{Quinn~Lanners}
\author[2]{Harsh~Parikh}
\author[3]{Alexander~Volfovsky}
\author[2]{Cynthia~Rudin}
\author[1]{David~Page}

\affil[1]{%
    Dept. of Biostatistics\\
    Duke University\\
    Durham, NC, USA.
}
\affil[2]{%
    Dept. of Computer Science\\
    Duke University\\
    Durham, NC, USA.
}
\affil[3]{%
    Dept. of Statistical Science\\
    Duke University\\
    Durham, NC, USA.
  }
  
\begin{document}
\maketitle
\begin{abstract}
Our goal is to produce methods for observational causal inference that are auditable, easy to troubleshoot, accurate for treatment effect estimation, and scalable to high-dimensional data. We describe a general framework called Model-to-Match that achieves these goals by (i) learning a distance metric via outcome modeling, (ii) creating matched groups using the distance metric, and (iii) using the matched groups to estimate treatment effects. Model-to-Match uses variable importance measurements to construct a distance metric, making it a flexible framework that can be adapted to various applications. Concentrating on the scalability of the problem in the number of potential confounders, we operationalize the Model-to-Match framework with LASSO. We derive performance guarantees for settings where LASSO outcome modeling consistently identifies all confounders (importantly \textit{without} requiring the linear model to be correctly specified). We also provide experimental results demonstrating the method's auditability, accuracy, and scalability as well as extensions to more general nonparametric outcome modeling.
\end{abstract}

\section{Introduction}
Matching methods are a popular approach to causal inference on observational data due to their conceptual simplicity. These methods aim to emulate randomized controlled trials by pairing similar treated and control units, thus allowing for treatment effect estimation \citep{stuart2010matching}. One significant benefit of using matching methods is their \textit{auditability} along with their accuracy. An auditable method allows domain experts to validate the estimation procedure, argue about the violation of key assumptions, and determine whether the analysis is trustworthy. Since causal analyses often depend on untestable assumptions, it is critical to determine whether all important confounders are accounted for, if data are processed correctly, and whether the treatment and control units in the matched groups are cohesive enough to be comparable \citep{pmlr-v162-parikh22a}. \cite{Parikh2022} and \cite{Ruoqi2021} showed that the audit of matched groups using external unstructured data is crucial in healthcare and social science scenarios. In high-stakes scenarios, audibility enables domain experts to make data-driven and \textit{trustworthy} decisions and policies.

We would ideally be able to match units that are exactly identical to one another except for treatment assignments \citep{rosenbaum1983central}. However, exact matches are almost impossible in high-dimensional settings with continuous covariates \citep{malts}. In such scenarios, we aim to create \textit{almost-exact matches} on important covariates. The question then becomes how to construct a distance metric between units that determines who should be in a unit's matched group. 
We want to learn a distance metric that provides \textit{accurate} causal estimates, ensures \textit{auditability} so we can evaluate and troubleshoot, and is \textit{scalable} to large observational datasets that might be used for high-stakes policy decisions.

We introduce the \textit{Model-to-Match} framework which uses variable importance from  prognostic score models to learn a distance metric. The framework has three steps. First, we use machine learning to estimate outcomes and use the measured variable importance to construct a distance metric. Second, we use the learned distance metric to match treatment and control units into matched groups. Third, we use the matched groups to estimate conditional average treatment effects (CATEs). Our research focuses on the first step in this framework, as learning a good distance metric is an essential but difficult step to ensure that matching yields accurate treatment estimates.

A special case of our framework is called \textit{LCM -- LASSO Coefficient Matching}. LCM has the characteristics we desire. It is able to accurately estimate treatment effects, creates almost-exact matched groups, and scales better than other comparable methods by orders of magnitude. LCM uses LASSO (Least Absolute Shrinkage and Selection Operator) coefficients to identify important variables and then uses K-nearest neighbors to construct matched groups. LCM benefits from both the efficiency of parametric models and the power of nonlinear modeling by leveraging a parametric method to learn which features to match on and then a nonparametric approach for treatment effect estimation. It is simple to implement yet works extremely well. We perform extensive empirical studies to compare LCM's performance with existing methods. Our results demonstrate that LCM can accurately and efficiently recover true treatment effects even in high-dimensional and non-linear setups without compromising auditability (Section~\ref{sec: results}). We further propose adaptations of our framework such as (a) metalearner LCM, (b) feature importance matching using decision trees, and (c) LCM-augmented-prognostic scores that work well in complex scenarios (Section~\ref{sec: extensions}).

\section{Background and assumptions}
We study the setting where every individual $i$ in the population $\mathcal{S}$ is assigned to one of the two treatments $T_i\in\{0,1\}$. Under the stable unit treatment value assumption (SUTVA), we define the potential outcomes of individual $i$ as $Y_i(0)$ and $Y_i(1)$. We consider an i.i.d sample of $n$ individuals, $\mathcal{S}_n$, where for each individual $i$ we observe a $p$-dimensional pre-treatment covariate vector $\X_i$, an assigned treatment $T_i$, and an observed outcome $Y_i = Y_i(1)T_i + Y_i(0)(1-T_i)$.

The individualized treatment effect is defined as $\tau_i := Y_i(1) - Y_i(0)$. Since we observe only one of the potential outcomes for each unit, $\tau_i$ is not observed for any of the units. We need to impute the missing potential outcomes to estimate the treatment effects of interest \citep{rubin2011causal}. 
In our setup, we are interested in identifying (a) conditional average treatment effects (CATEs) \(\tau(\x) := \mathbb{E}\left[\tau_i \mid \X_i=\x\right]\) for all \(\x \in Dom(\X)\), and (b) the average treatment effect (ATE) \(\tau := \mathbb{E}\left[\tau_i\right]\).

 In observational data (where the treatments are not randomized), the treatment choice and potential outcomes can depend on common variables, which are referred to as confounders. In our setup, we assume that the set of confounders is a subset of the set of pre-treatment covariates, and potential outcomes and treatment assignment are conditionally independent given $\X$: $(Y_i(1),Y_i(0)) \perp T_i \mid \X_i$. This is referred to as conditional ignorability \citep{rubin1974estimating}. Lastly, we assume that the probability of a unit receiving treatment $t$ is bounded away from $1$ and $0$: $0 < P(T_i = t\mid \X_i = \x) < 1$. This is referred to as the positivity assumption. Combining the positivity and conditional ignorability assumptions, adjusting for pre-treatment covariates ($\X$) is sufficient to identify CATEs and ATE. 

\paragraph{Matching Methods.} Matching methods use a distance-metric, $d_{\mathcal{M}}$, on $\X$ to group similar units with different treatment assignments in order to estimate the causal effects of treatment $T$ on outcome $Y$. The most popular matching techniques are \textit{propensity score matching} (PSM) \citep{rosenbaum1983central} and \textit{prognostic score matching} (PGM) \citep{hansen2008prognostic}. These techniques project the data to a lower dimensional propensity or prognostic score, which are then used for matching. These projections can be sensitive to modeling choices that affect the accuracy of the treatment effect estimates \citep{modelmisspecificaton}. Further, the units within a matched group can be far from each other in covariate space -- i.e., the matched groups are generally not auditable \citep{malts}. To date, the only observational causal inference techniques that attempt to optimize accuracy while maintaining auditability are those stemming from the almost-matching-exactly (AME) framework, namely the optimal matching (optMatch) \citep{Ruoqi2021, pmlr-v54-kallus17a}, genetic matching (GenMatch) \citep{genmatch}, FLAME/DAME \citep{wang2017flame, DiengEtAl2018}, MALTS \citep{malts, parikh2019application, Parikh2022} and AHB \citep{morucci2020adaptive} algorithms. FLAME/DAME can scale to extremely large datasets but handles only categorical variables. GenMatch, MALTS, and AHB can also handle both continuous and categorical variables but do not scale as well, thereby limiting their usefulness (see Figure~\ref{fig:scaling-runtime} in Section~\ref{sec: results}). What we develop is a method that yields accurate treatment effect estimates and is auditable like MALTS but can scale to much larger datasets and run at a fraction of the time.

Formally, for a unit $i$, the K-nearest neighbors of units with treatment $t'$ and the corresponding matched group $\text{MG}_{d_{\mathcal{M}}}(\X_i)$ are defined as 
\begin{equation*}
    \begin{gathered}
        \text{KNN}_{d_{\mathcal{M}}}(\X_i,t') := \\ 
                \left\{k:\sum_{j\in \mathcal{S}^{(t')}_n} \mathbbm{1}
                \begin{bmatrix}
                d_{\mathcal{M}}(\X_i,\X_j)\;\;\\
                \;\; < d_{\mathcal{M}}(\X_i,\X_k)
                \end{bmatrix}
                < K\right\},    
    \end{gathered}
\end{equation*}
\begin{equation*}
    \text{MG}_{d_{\mathcal{M}}}(\X_i):=\bigcup_{t'\in\{0,1\}} \text{KNN}_{d_{\mathcal{M}}}\left(\X_i,t'\right),    
\end{equation*}
where $\mathcal{S}^{(t')}_n:= \left\{j : T_j = t' \right\}$ represents the set of units whose treatment assignment is $t'$.
Match groups can then be used to estimate potential outcomes, 
    \(\widehat{Y}_i(t') = \psi\left(\text{KNN}_{d_{\mathcal{M}}}(\X_i,t')\right)\),
where $\psi$ is a function of the outcomes of the K-nearest neighbors (e.g. arithmetic mean). As we will see, a high quality distance metric is key to creating accurate estimates. A good distance metric can lead to interpretable matched groups and accurate treatment effect estimates; a poor distance metric leads to neither.

\paragraph{Non-matching Methods.} There are a number of non-matching frameworks that can estimate conditional average treatment effects. Regression methods, particularly doubly robust regression methods, are often used to estimate CATEs \citep{doublyrobust}. However, their performance is highly sensitive to model misspecification, requiring either the propensity or outcome model to be correctly specified. Machine learning methods are also popular for estimating CATEs. The most commonly used machine learning methods include Bayesian additive regression trees (BART) \citep{bart}, double machine learning \citep{chernozhukov2017doubledebiased}, and generalized random forests \citep{grf-article}. While these methods can accurately estimate CATEs, they are often significantly less interpretable than matching methods and are not auditable. Additionally, previous almost-matching-exactly literature has shown that AME methods achieve similar CATE estimation accuracy to machine learning approaches while maintaining auditability \cite{malts, morucci2020adaptive, wang2017flame}. For these reasons, in this paper we focus on comparing LCM to other matching methods and AME methods in particular. We include an experiment comparing LCM to machine learning methods on a high-dimensional quadratic dataset in the Supplementary Material.

\section{Model-to-Match Framework}\label{sec: framework}
We propose a framework, called \textit{Model-to-Match}, that focuses on combining prognostic score modeling with distance metric learning for almost-exact matching. 
Our framework is divided into three steps: (i) learning the weight matrix $\mathcal{M}$ of a distance metric $d_{\mathcal{M}}$ using a machine learning model, (ii) creating matched groups using the learned $d_{\mathcal{M}}$, and (iii) estimating treatment effects using the matched groups. 

In our framework, we restrict ourselves to binary and continuous pre-treatment covariates. As such, all categorical covariates are dummified in the data preprocessing steps. We let $p$ indicate the dimensionality of the final covariate space after preprocessing. This facilitates the use of more feature-importance methods (such as LASSO) and allows the feature space to be more finely weighted.

We choose our distance metric, $d_\mathcal{M}$, such that for any $\X_1$ and $\X_2$, $d_{\mathcal{M}}(\X_1, \X_2) = \| \mathcal{M} \X_{1}- \mathcal{M} \X_{2}\|_{m}$. $\mathcal{M}$ is a $p\times p$ matrix and the $m$ in $\|\cdot\|_m$ is flexible and can be any positive integer.

To learn a distance metric in our Model-to-Match framework we first train two machine learning models $f^{(0)}$ and $f^{(1)}$, such that for any $i \in \mathcal{S}_n$, $\widehat{Y}_i(t') = f^{(t')}(\X_i)$. 
For each $j\in\{1,2,\dots,p\}$ we then calculate $\theta_j$, the importance of covariate $X_{\cdot, j}$ to $f^{(0)}$ and $f^{(1)}$.

\textit{Variable Importance Example 1:} If the $f$'s are linear estimators, such as LASSO or Ridge where $f^{(t)}_{\bm{\beta}^{(t)}} = \X\bm{\beta}^{(t)}$, then $\theta_j$ can be $\sum\limits_{t'\in\{0,1\}} \frac{|\beta^{(t')}_j|}{\|\bm{\beta}^{(t')}\|_1}$.

\textit{Variable Importance Example 2:} If the $f$'s are decision trees then $\theta_j$ can be measured via Gini importance, feature permutation importance, or a similar feature importance metric.

\textit{Variable Importance Example 3:} For $f$'s from backward elimination with ordinary least squares, $\theta_j$ can be equal to the drop in $R^2$ when the $j$-th feature $X_{\cdot,j}$ is removed. 

\textit{Variable Importance Example 4:} For any generic model class, $\theta_j$ can be measured via subtractive model reliance, which measures the change in the loss of a model when a covariate is perturbed \citep{fisher2019all}. 

We then set all the diagonal entries, $\mathcal{M}_{j,j}$, in the distance metric $\mathcal{M}$ to be equal to $|\theta_j|$ and all the non-diagonal entries in $\mathcal{M}$ to zero. By constructing $\mathcal{M}$ in this way we can interpret each weight, $\mathcal{M}_{j,j}$, as the relative feature importance of covariate $X_{\cdot,j}$.

We are interested in having an $\mathcal{M}$ that is sparse so that we only match on the important covariates. Further, we want the estimation of $f$'s to be scalable in both the number of samples and the number of covariates. Keeping these requirements in mind, we use $\ell_1$-regularized regression, i.e., LASSO, as the modeling method of choice for the majority of this paper. However, our framework is general and can be applied to any supervised model class. For example, we discuss using shallow regression trees to model the $f$'s in Section~\ref{sec: extensions}. In practice, LASSO performs well for this step of the framework.

\section{Linear Coefficient Matching} \label{sec: method}

In this section, we operationalize the \textit{Model-to-Match} framework using LASSO \citep{Tibshirani1996} as the machine learning algorithm for learning the distance metric and refer to this as LASSO Coefficient Matching (LCM). As in the example in Section~\ref{sec: framework}, we use scaled absolute values of LASSO's coefficients as the diagonal entries for an $\mathcal{M}^*$. Since LASSO's coefficients are sparse, the entries of $\mathcal{M}^*$ will be sparse.
This creates a distance metric $d_{\mathcal{M}^*}$ that prioritizes tighter matches on a small number of important covariates, leading to faster runtimes and facilitating matched groups that are close in important covariates.

We perform \textit{honest} causal estimation for a given observed dataset $\mathcal{S}_n$. Broadly, honest causal estimation means that we do not use the same data to learn about the control variables as we do for inference \citep{Ratkovic2019RehabilitatingTR}. We achieve honesty by dividing the data into two disjoint subsets: $\mathcal{S}_{n, tr}$ and $\mathcal{S}_{n, est}$. In Step (i), we use $\mathcal{S}_{n, tr}$ to estimate $\bm{\beta}$'s and, by consequence, learn $d_{\mathcal{M}^*}$. Algorithm~\ref{alg:lcm} describes our training step to learn $\mathcal{M}^*$ using LASSO. 
In Step (ii), we then perform matching with replacement using $d_{\mathcal{M}^*}$ to get matched groups, $\text{MG}_{d_{\mathcal{M}^*}}(\X_i)$, for each unit $i \in \mathcal{S}_{n, est}
$. In Step (iii), we use $\text{MG}_{d_{\mathcal{M}^*}}(\X_i)$ to estimate the CATE for $\X=\X_i 
$ as $\widehat{\tau}(\X_i) = \widehat{Y}_i(1) - \widehat{Y}_i(0)$ where 
\begin{equation*}
    \widehat{Y}_i(t') = \frac{\sum_{k \in \text{MG}_{d_{\mathcal{M}^{*}}}(\X_i)} \mathbbm{1}[T_k = t'] Y_k}{\sum_{k \in \text{MG}_{d_{\mathcal{M}^*}}(\X_i)} \mathbbm{1}[T_k = t']}.
\end{equation*}

\begin{algorithm} \label{alg}
{\DontPrintSemicolon 
\KwData{Dataset $\mathcal{S}_{n, tr}$}
\KwResult{Distance metric $\mathcal{M}^{*}$}
\Begin{ 
    $W \leftarrow [0,..., 0]\in\mathbb{R}^p$\;
    \textrm{\rm (Loop over treatment possibilities.)}\;
    \For{$t'$ in $\{0,1\}$ }{ 
        \textrm{(Find units that have treatment }$t'$\textrm{.)}\;    
        \(\mathcal{S}_{n, tr}^{(t')} \leftarrow \{ i\in \mathcal{S}_{n, tr} : T_i = t' \}\)\;
        \textrm{(Run LASSO to get coefficients.)}\;
        \(\hat{\bm{\beta}}^{(t')} \leftarrow \min_{\bm{\beta} \in \mathbb{R}^{p}} \lambda\| \bm{\beta} \|_{1} + \sum\limits_{i \in \mathcal{S}_{n, tr}^{(t')}} \left( Y_i - \X_i \bm{\beta} \right)^{2} \)\;
        \textrm{(Average the element wise absolute values of the coefficients across treatment and control.)}\;
        \For{\(l\) in \(\{1,...,p\}\)}{
            \(W_l \leftarrow W_l + \frac{|\hat{\beta}^{(t')}_l|}{\|\hat{\bm{\beta}}^{(t')}\|_1}\)\;
        }
    }
    \textrm{(Coefficients used as stretches in distance metric.)}\;
    \(\mathcal{M}^{*} \leftarrow \textbf{0}_{p\times p}\)\;  
    \For{\(l\) in \(\{1,...,p\}\)}{
        \(\mathcal{M}^{*}_{l,l} \leftarrow \frac{1}{2} \; W_l\) 
    }
}}
\caption{Algorithm to estimate $\mathcal{M}^{*}$ using LASSO}
\label{alg:lcm}
\end{algorithm}
Since we perform honest causal inference where we do not use the same data to learn $d_{\mathcal{M}^*}$ as we do for estimating CATEs, our method performs $\eta$-fold cross-fitting by swapping the training set each time. This is similar to the strategy used in \cite{chernozhukov2018double} and enables the estimation of CATEs for all $i \in \mathcal{S}_n$. Because LASSO does not need many observations to fit the data well, we use only one of the $\eta$ splits as the training set and the data in the remaining $(1-\eta)$ splits as the estimation set. Using a smaller amount of data in the learning step allows us to create match groups with a larger portion of the data. Because the nearest neighbor-based estimation in Step (iii) is local and non-parametric, more data will improve the quality of matched groups and the accuracy of the CATEs.

For matching we employ the Manhattan distance to align with the additive linear form and $\|\cdot\|_1$ regularization of LASSO. In particular, for all $i,j \in \mathcal{S}_{n, est}$,
$d_{\mathcal{M}^{*}}(\X_i, \X_j) = \sum\limits_{l=1}^p \mathcal{M}^{*}_{l,l}|\X_{i,l} - \X_{j,l}|$. Our method has three hyperparameters: $\eta$, $\lambda$, and $K$. We learn $\lambda$ using cross-validation in our training in Step (i). The number of nearest neighbors, $K$, and the number of splits for cross-fitting, $\eta$, can be chosen through cross-validation or set manually. 

\section{Theoretical Results} \label{sec: theory}

Here, we prove optimality properties of using LASSO to learn our distance metric. We then show under what conditions LCM guarantees consistency in CATE estimation. Proofs are included in the Supplementary Materials.

Theorem~\ref{thm:motivation} motivates LCM. It shows that if the potential outcomes are linear in the predictors then using the absolute values of the coefficients in these models as the stretches in a distance metric guarantees that as the distance between two units decreases, their expected outcomes become closer.

\begin{theorem}\label{thm:motivation}[Closeness in $\X$ implies closeness in $Y$].
Consider a $p$-dimensional covariate space where for $t' \in \{0,1\}$, $f^{(t')}(\X_i) = \mathbb{E}[Y_i | \X = \X_i, T = t' ] = \X_i\bm{\beta}^{(t')}$. Construct $\mathcal{M}\in\mathbb{R}^{p\times p}$ where for all $l,r \in \{1,...,p\}$ $\mathcal{M}_{l,l} = |\beta^{(t')}_l|$ and for $l\neq r$ $\mathcal{M}_{l,r} = 0$. Then, $\forall i, j$, we have that $d_{\mathcal{M}}(\X_i, \X_j) \geq \left|f^{(t')}(\X_i) - f^{(t')}(\X_j) \right|$.
\end{theorem}

From here, we define a diagonal Mahalanobis distance matrix as any $\widetilde{\mathcal{M}}\in\mathbb{R}^{p\times p}$ that is diagonal (for all $l,r \in \{1,...,p\}$, $l\neq r$, $\widetilde{\mathcal{M}}_{l,r} = 0$) and has non-negative entries ($\widetilde{\mathcal{M}}_{l,l} \geq 0$). We show in Theorem~\ref{thm:supp} that the $\mathcal{M}$ from Theorem~\ref{thm:motivation} is the optimal stretch matrix, compared to any other equally scaled diagonal Mahlanobis distance matrix, in regards to the maximum absolute difference in expected outcomes.

\begin{theorem}\label{thm:supp}[Optimality of $\mathcal{M}$]
    Using the setup of Theorem~\ref{thm:motivation}, let $\supp(\X) = \mathbb{R}^p$.
    Consider an arbitrary diagonal Mahalanobis distance matrix $\widetilde{\mathcal{M}}\in\mathbb{R}^{p\times p}$ where $\sum\limits_{l=1}^p|\widetilde{\mathcal{M}}_{l,l}| = \sum\limits_{l=1}^p |\bm{\beta}^{(t')}_l|$ and $\widetilde{\mathcal{M}}_{l,l} > 0$ when $|\beta^{(t')}_l| > 0$.
    For some $\epsilon \geq 0$ and $\X_1\in\mathbb{R}^p$, define $S_{\widetilde{\mathcal{M}}, \epsilon}(\X_1) := \{\X_2 : \X_2\in\mathbb{R}^p, d_{\widetilde{\mathcal{M}}}(\X_1, \X_2) = \epsilon\}$. Then, 
   \begin{align*}
        \sup\limits_{\X_2\in S_{\mathcal{M}, \epsilon}(\X_1)}|f^{(t')}(\X_1)- f^{(t')}(\X_2)| 
        \leq\\
        \sup\limits_{\X_3\in S_{\widetilde{\mathcal{M}}, \epsilon}(\X_1)}|f^{(t')}(\X_1)- f^{(t')}(\X_3)|.
    \end{align*}
\end{theorem}

These results show how a linear outcome model induces a meaningful distance metric for causal inference. The following theorem states that when we do not know the true value of the coefficients (and more generally when the model is non-linear but LASSO still recovers its support), we can employ the LCM procedure of Section~\ref{sec: method} to generate a distance metric that yields consistent estimates of CATEs. This theorem uses the notion of variable importance, as discussed in Section \ref{sec: framework}.

\begin{theorem}\label{thm:consistency}[Consistency of LCM]
     For $t'\in\{0,1\}$, let $f^{(t')}(\X_i) = \mathbb{E}[Y_i | \X = \X_i, T = t' ] $. Let $f^{(t')}$ be Lipschitz continuous and,
     \begin{equation*}
        \supp\left(f^{(t')}\right) :=  \left\{j : 
             \textrm{importance of }\X_{\cdot, j}\textrm{ in } f^{(t')}\textrm{ is } > 0\right\}.  
     \end{equation*}
     Denote $d_{\mathcal{M}^*}$ as the distance metric learned by LCM in Section~\ref{sec: method} and let $\Gamma\left(\mathcal{M}^*\right) = \{j : \mathcal{M}^*_{j,j} > 0\}$. LCM is consistent for CATE estimation if $\supp\left(f^{(0)}\right) \bigcup \supp\left(f^{(1)}\right) \subseteq \Gamma\left(\mathcal{M}^*\right)$.
\end{theorem}

This result follows from LASSO and its adaptations' ability to estimate sparse coefficient vectors in high dimensions, even when $n < p$ \citep{Meinshausen2009, Zhou2010, Wasserman2009, Meinshausen2006}. LASSO also exhibits consistency for feature selection in some nonlinear settings \citep{Zhang2016a}. 

\section{Experimental Results}\label{sec: results}

Our  experiments focus on factors crucial in high-stakes causal inference.
\textbf{(i) Accuracy and Auditability}: We compare LCM's  matched groups to PGM's and highlight the importance of auditability. 
\textbf{(ii) Nonlinear Outcomes}: We study if LCM is sensitive to model misspecification and compare our results to linear PGM (which uses the same underlying prognostic model as LCM). 
\textbf{(iii) Scalability}: We compare LCM to existing AME algorithms in both runtime and estimation performance as both the number of observations and the number of features increase.

\subsection{Accuracy and Auditability}\label{sec:results-auditability}
Matching enables us to investigate whether a CATE is estimated in a trustworthy manner by \textit{auditing} the quality of the matched groups. We now highlight how LCM produces accurate estimates while matching tightly on important covariates.
We work with the ACIC 2018 Atlantic Causal Inference Conference semi-synthetic dataset \citep{Carvalho2019}, which is based on data from the National Study of Learning Mindsets randomized trial \citep{Yeager2021}. The dataset contains 10,000 students across 76 schools. There are four categorical student-level covariates and one categorical and five continuous school-level covariates. \citet{Carvalho2019} constructed this semi-synthetic dataset by drawing covariates from the real experiment and then synthetically generating treatment assignments and outcomes. Details can be found in \cite{Carvalho2019}.

We ran our method alongside linear PGM, computed using LASSO, and nonparametric PGM, computed using gradient boosted trees. All three methods recover ATE estimates that are close to the true value of 0.24 -- \textbf{LCM}: 0.249, \textbf{Linear PGM}: 0.251, and \textbf{Nonparametric PGM}: 0.260, which are also in line with the estimates of other interpretable and uninterpretable methods described in \cite{Carvalho2019}. 

\begin{figure}
\centering
\includegraphics[width=\linewidth]{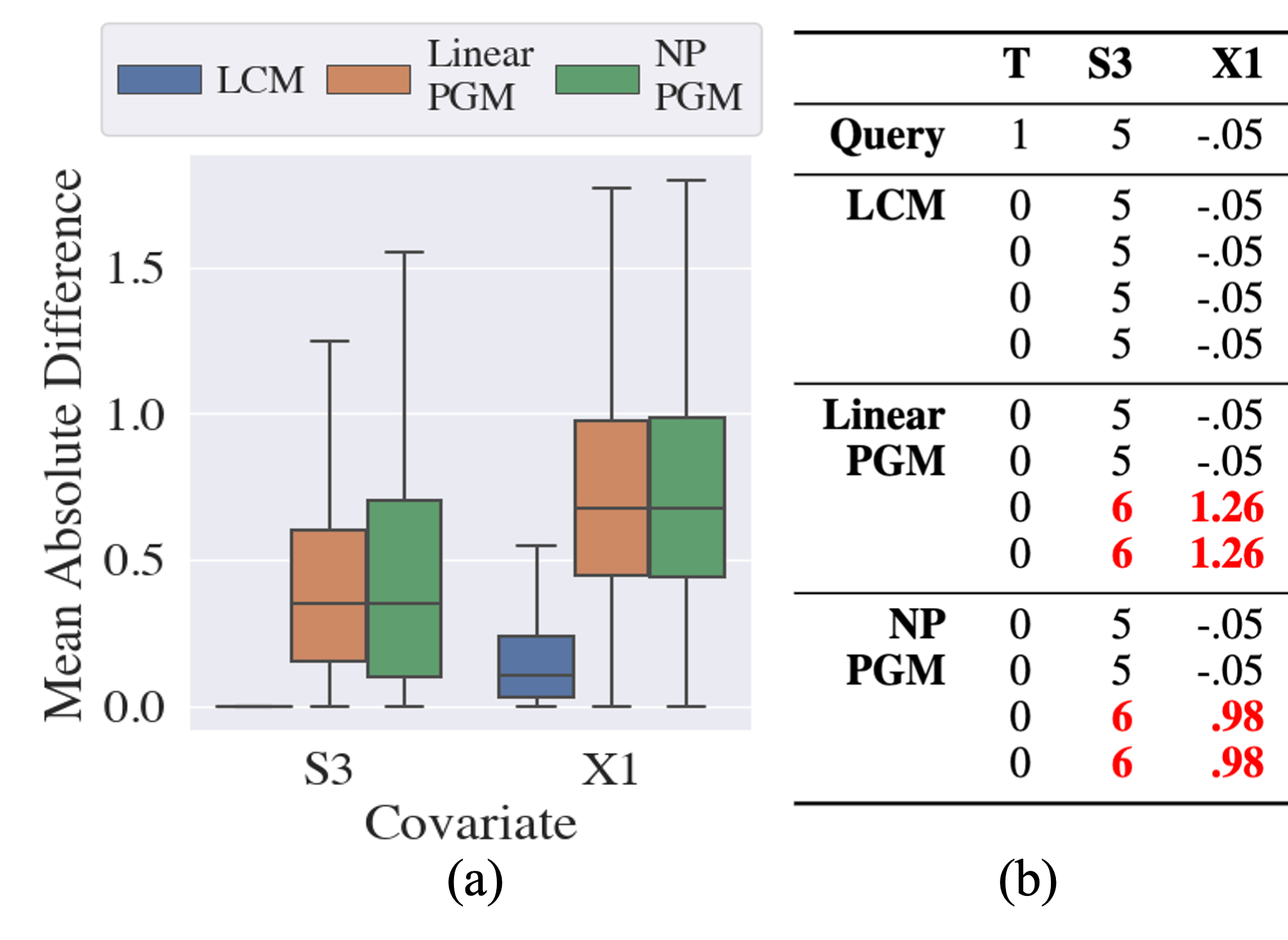}
  \caption{Closeness in important covariates for matched groups produced by LCM, linear PGM, and nonparametric (NP) PGM. \textit{(a)} shows the mean absolute difference between a query unit and its matched group's covariate values. Smaller values imply better and tighter matches. \textit{(b)} shows, for a random sample, the four nearest neighbors of opposite treatment under LCM, linear PGM, and NP PGM. In \textit{(b)}, the text in \textcolor{red}{red} indicates values that are far from the query unit's value. $\text{S3}$ indicates the self-reported prior achievements of students and is important for selection into treatment, and $\text{X1}$ indicates school-level average mindset score of the students and is an effect modifier.
  }
  \label{fig:schools-mg-diff}
\end{figure}

While all three methods accurately estimate the ATE, \textit{only LCM matches almost exactly on important covariates}. We compare how tightly LCM, linear PGM, and nonparametric PGM fit on a covariate that is identified as important for selection into treatment (S3) and one that is an effect modifier (X1). 
Figure \ref{fig:schools-mg-diff} shows that LCM matches tighter on important covariates than PGM. In this way, LCM more closely emulates exact-matching and results in more intuitive and auditable match groups. The fact that LCM accurately estimates the treatment effect and matches so tightly on these important covariates increases the trust we have in our conclusions. We expand on these findings and show that LCM matches tighter across all the effect modifiers in the Supplementary Material.

\subsection{Nonlinear Outcomes}\label{sec:results-nonlinear}
We have shown that linear prognostic score matches are not tight on important covariates, leading to unintuitive matched groups. However, LASSO estimated prognostic scores are more interpretable than scores estimated with gradient boosted trees. This interpretability comes at a cost: the performance of linear PGM heavily depends on the linearity of the underlying data generation process. LCM is more robust to nonlinear data because its LASSO component is used \textit{only to determine the relative weight of features} in the distance metric (not to model the outcome with a linear combination of the covariates).

We compare CATE estimation accuracy of LCM and linear PGM on two synthetically generated datasets where the outcome is a non-linear function of the covariates. We call these datasets \textbf{Sine} and \textbf{Exponential} to align with their underlying potential outcome functions. We generate 5000 samples and 100 covariates for each dataset. For \textbf{Sine}, the outcome function is
\begin{equation*}
    Y_i = \sin(X_{i,1}) - T_i \sin(X_{i,2}).
\end{equation*}
Whereas, for \textbf{Exponential}, the outcome function is
\begin{equation*}
    Y_i = 2e^{X_{i,1}} - \sum_{j=2}^3 e^{X_{i,j}} + T_i e^{X_{i,4}}.
\end{equation*}
We outline the specific details of the data generation processes in the Supplementary Material. Figure \ref{fig:lcm-vs-lin-pgm} shows that LCM is more robust to nonlinear outcome functions than linear PGM.

Again, the superior performance of LCM is unsurprising because it performs nonlinear estimation in Step (iii), using the linear LASSO method in Step (i) only to pinpoint important covariates upon which nonlinear estimation can be successfully performed.

\begin{figure}
\centering
\includegraphics[width=\linewidth]{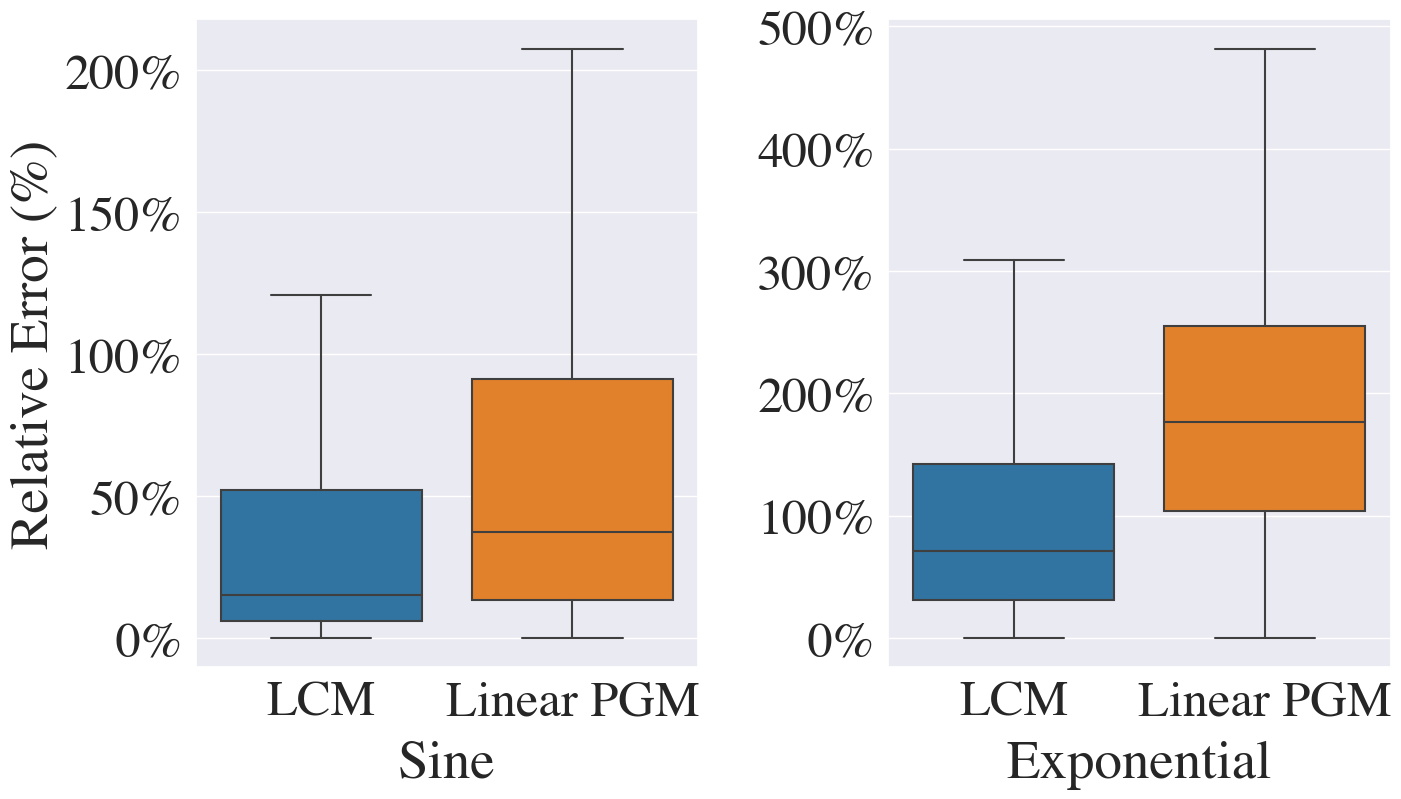}
  \caption{CATE estimation accuracy of LCM and Linear PGM on nonlinear synthetically generated datasets \textbf{Sine} and \textbf{Exponential}. The y-axis is the absolute CATE estimation error relative to the true ATE.}
  \label{fig:lcm-vs-lin-pgm}
\end{figure}

\subsection{Scalability}\label{sec:results-scalability}
Existing almost-matching-exactly methods learn covariate weights and/or create match groups through computationally expensive and data hungry optimization algorithms. In this section we compare LCM to MALTS \citep{malts}, GenMatch \citep{genmatch}, and AHB \citep{morucci2020adaptive} in regards to scalability in runtime and CATE estimation accuracy. We omit FLAME/DAME \citep{wang2017flame,DiengEtAl2018} from this comparison since it can only handle discrete covariates.

\begin{figure}
\centering
\includegraphics[width=\linewidth]{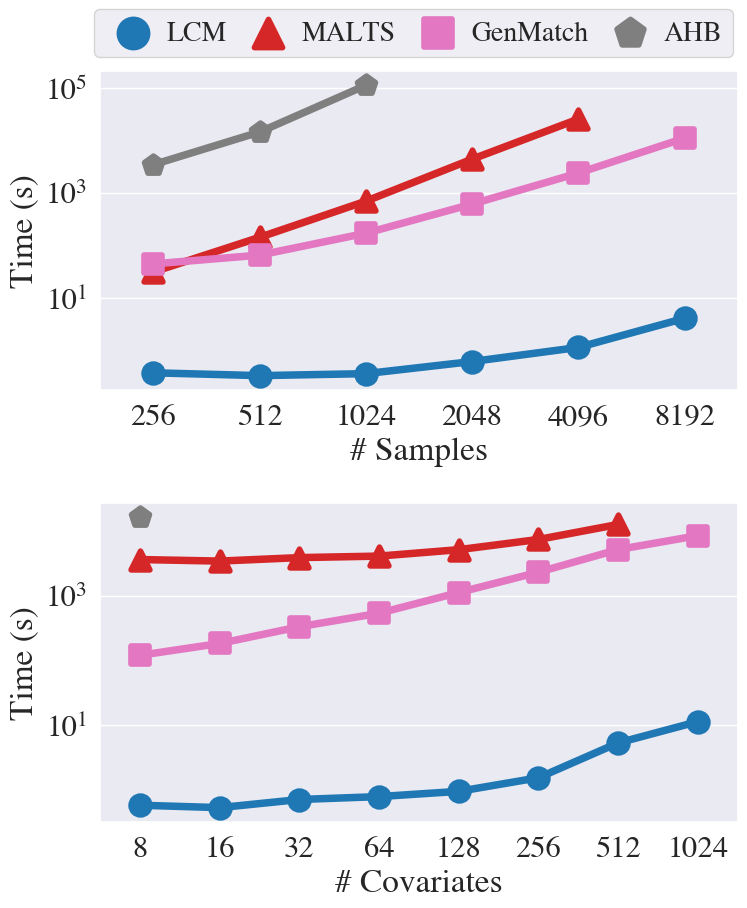}
  \caption{Scalability in $n$ and $p$ for LCM, MALTS, GenMatch, and AHB.}
  \label{fig:scaling-runtime}
\end{figure}

We generate synthetic datasets of various sizes from the quadratic data generation process described in \cite{malts} and the Supplementary Materials. We first measure the runtime scalability of LCM, MALTS, GenMatch, and AHB with respect to the number of samples, $n$, and number of covariates, \(p\). To measure scaling runtime in \(n\), we keep the number of covariates constant at 64 and increase the number of samples from 256 to 8192. To measure scaling in \(p\), we set the number of samples to be 2048 and vary the number of covariates from 8 to 1024. The Supplementary Materials contain further information on how runtimes were measured. Figure \ref{fig:scaling-runtime} shows the runtimes for each of the AME algorithms on these various dataset sizes, highlighting the multiple-order-of-magnitude runtime disparity between LCM and other AME methods. MALTS ran out of memory (16GB RAM) for the largest dataset in each plot. We stopped increasing the dataset sizes for AHB when its runtime surpassed the longest measured runtime of all other methods.

\begin{figure}
\centering
\includegraphics[width=\linewidth]{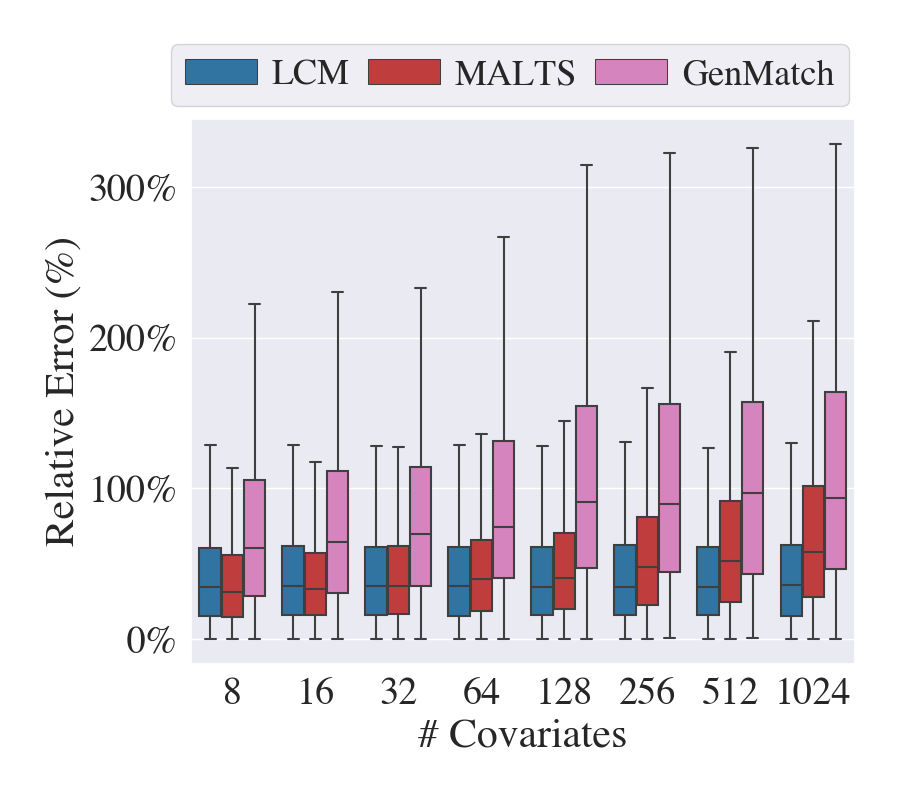}
  \caption{Absolute CATE estimation error relative to the true ATE for LCM, MALTS, and GenMatch as the number of covariates increases.}
  \label{fig:scaling-accuracy}
\end{figure}

As discussed in Section \ref{sec: method}, LASSO is capable of recovering sparse \(\bm{\beta}\)s and important features in high dimensional settings. Naturally, LCM also excels at producing accurate CATE estimates as the number of irrelevant covariates grows. Figure \ref{fig:scaling-accuracy} shows how LCM is robust to added noise as the number of unimportant covariates grows -- unlike MALTS and GenMatch, which struggle to learn an accurate distance metric as the dimensionality of the covariate space increases. Here, we keep the number of important covariates equal to 8.

\section{Model-to-Match Adaptations}\label{sec: extensions}
In this section, we propose three adaptations of the Model-to-Match framework that extend LCM. The first approach uses a metalearner variant of LCM and shows improvement in CATE estimation in certain settings. The second adaptation proposes the use of a tree-based outcome modeling approach in place of LASSO. The third adaptation combines prognostic score matching with LCM to yield accurate CATEs and tight match groups.
\subsection{Metalearner LCM}\label{sec: metalearner}
Metalearners leverage powerful regression tools for estimating heterogenous treatment effects \citep{Knzel2019}. LASSO Coefficient Matching can be adapted to run similar to the T-learner outlined in \cite{Knzel2019} by learning separate distance metrics for control and treated units. The metalearner adaptation of LCM is advantageous when certain covariates have vastly different effects on the outcome depending on if a sample received treatment or not. 

\begin{figure}
\centering
  \includegraphics[width=\linewidth]{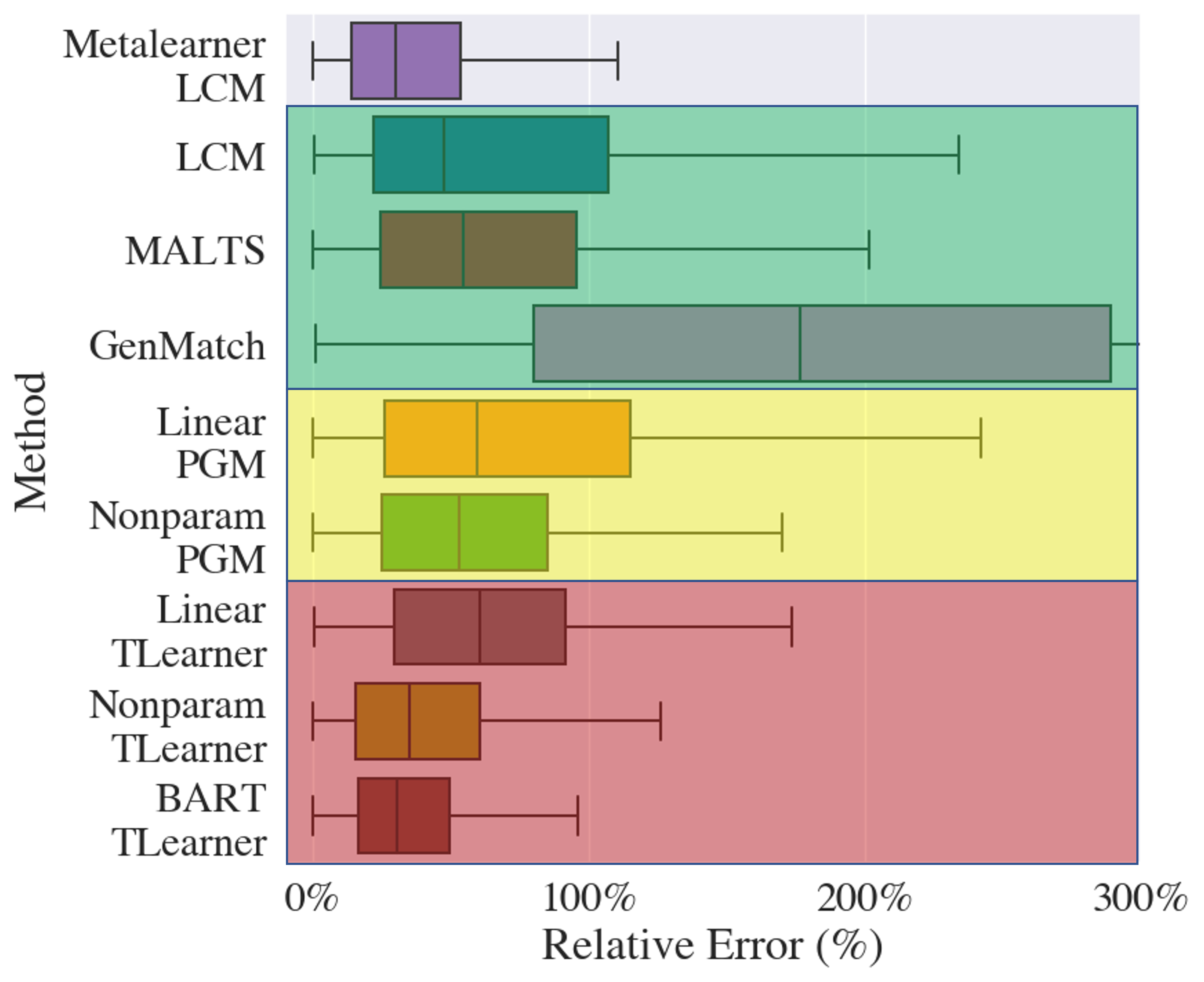}
  \caption{Absolute CATE estimation error relative to the true ATE for various methods on the \textbf{Sine} data generation process. The transparent boxes separate the methods into different categories. \textbf{Green}: Almost exact matching methods. \textbf{Yellow}: Other matching methods. \textbf{Red}: TLearner methods.}
  \label{fig:metalearner-error}
\end{figure}

\begin{figure}
\centering
  \includegraphics[width=0.8\linewidth]{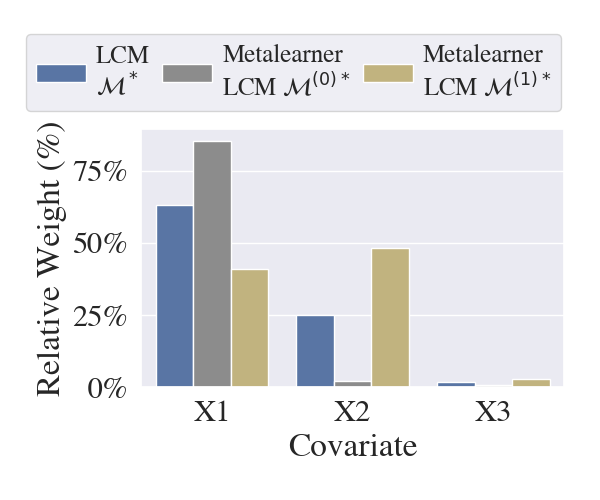}
  \caption{Relative covariate weights averaged over the $\eta$-folds for LCM $\mathcal{M}^*$, Metalearner LCM $\mathcal{M}^{(0)*}$, and Metalearner LCM $\mathcal{M}^{(0)*}$. This shows that the Metalearner LCM's distance metrics are different between treatment and control groups.}
  \label{fig:metalearner-weights}
\end{figure}

For Metalearner LCM, we learn a separate distance metric, $d_{\mathcal{M}^{(t')*}}$, for each $t'\in\{0,1\}$. Specifically, for $l,r\in\{1,\ldots,p\}$ we set $\mathcal{M}^{(t')*}_{l,l} = |\hat{\beta}^{(t')}_l|\frac{1}{\|\hat{\bm{\beta}}^{(t')}\|_1}$, where $\hat{\bm{\beta}}^{(t')} = \min_{\bm{\beta} \in \mathbb{R}^{p}} \lambda\| \bm{\beta} \|_{1} + \sum_{i \in \mathcal{S}_{n, tr}^{(t')}} \left( Y_i - \X_i \bm{\beta} \right)^{2}$, and $\mathcal{M}^{(t')^*}_{l,r} = 0$ when $l\neq r$. In Step (ii), for each unit $i \in \mathcal{S}_{n, est}$, we find K-nearest neighbors with replacement using the corresponding distance metric in each treatment arm.

To illustrate the advantage of the Metalearner LCM, we consider the same \textbf{Sine} data generation process used in Section~\ref{sec:results-nonlinear}. In \textbf{Sine}, covariate \(X_{i,1}\) is important to the outcome under both treatment regimes (\(Y_i(0)\) and \(Y_i(1)\)) while covariate \(X_{i,2}\) is only relevant to the outcome under treatment (\(Y_i(1)\)). We generate 500 samples and 10 covariates. We compare LCM to the previously used matching methods along with linear and nonparametric T-Learners. Figure \ref{fig:metalearner-error} shows estimated CATE errors for each method. Metalearner LCM improves upon LCM, which already outperforms other matching methods, and is comparable to T-Learners.

Figure \ref{fig:metalearner-weights} shows how Metalearner LCM stretches  the control and treatment response surfaces differently, whereas regular LCM learns a global metric that is a linear combination of the two treatment spaces. The Metalearner variant is more suitable for problems in which accurate CATE estimation is more important than emulating a randomized experiment.

\subsection{Feature Importance Matching}\label{sec:feature-imp}
LASSO can often find the important features, even if the true data generation process is nonlinear \citep{Zhang2016a}. However, in cases where it cannot, we can use any nonlinear method (decision tree, random forest, BART, AdaBoost, etc.) from which we can extract a measure of feature importance. These feature importance values can be used in place of LASSO coefficients in Algorithm~\ref{alg:lcm} as weights for matching.

We demonstrate this using shallow decisions trees as the model and Gini importance as the feature importance measure \citep{menze2009comparison}. We use a shallow decision tree to promote sparsity and to account for nonlinearities in the outcome space.
We generate a dataset with 1000 samples and 10 covariates where only the first covariate is important: $Y_i(0) = X_{i,1}^2 + \epsilon_{i,y}$ and $Y_i(1) = X_{i,1}^2 + 10 + \epsilon_{i,y}$. A linear approach will not find this important covariate because it is symmetric around 0. The full data generation process is outlined in the Supplementary Material. 
Figure \ref{fig:lcm-vs-tree} shows that in this setting, the tree-based method creates more accurate CATE estimates than the LASSO method (LCM).

\begin{figure}
\centering
  \includegraphics[width=0.7\linewidth]{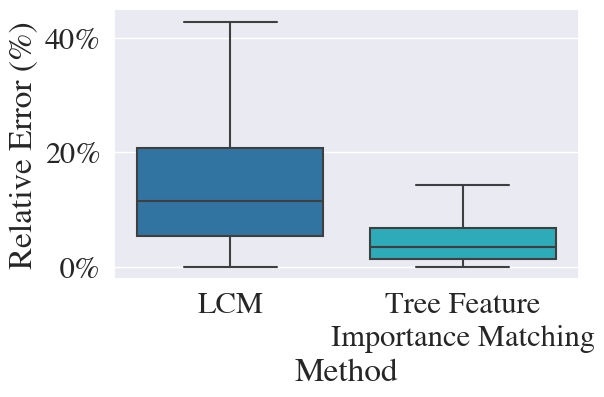}
  \caption{Absolute CATE estimation error relative to the true ATE for LCM vs$.$ the Model-to-Match framework with classification and regression decision tree (CART) as the model and Gini feature importance as the feature importance measure.}
  \label{fig:lcm-vs-tree}
\end{figure}

\subsection{LCM-Augmented-PGM}\label{sec:lcm-aug-pgm}
As shown in Section~\ref{sec:results-auditability}, LCM produces tighter matched groups on important covariates than linear and non-parametric PGM. However, PGM sometimes can estimate CATEs more accurately while not producing tight matched groups. This might occur either when the parametric prognostic model is correctly specified or when there is a strong non-linear effect that a non-parametric prognostic score can model accurately. In such situations, we propose augmenting PGM with LCM to guarantee tight matches and accurate CATE estimates. Our LCM-augmented-PGM (LAP) is a two stage procedure. In the first stage, we match using prognostic scores and create large matched groups. In the second stage, we match using the distance metric learned via LCM inside each PGM matched group. The first stage leverages the flexibility of outcome modeling and the second stage ensures tight matching on important covariates.

\begin{figure}
    \centering
    \includegraphics[width=0.4\textwidth]{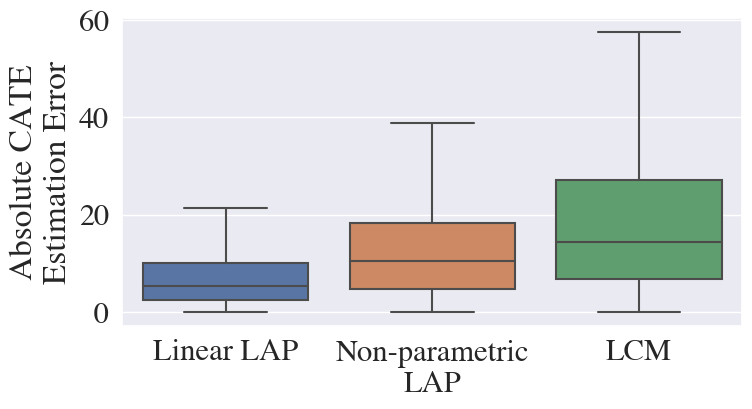}
    \caption{Absolute CATE estimation error for linear LAP (blue), non-parametric LAP (orange) and LCM (green). }
    \label{fig:lap_a}
\end{figure}

\begin{figure}
\centering
\includegraphics[width=0.4\textwidth]{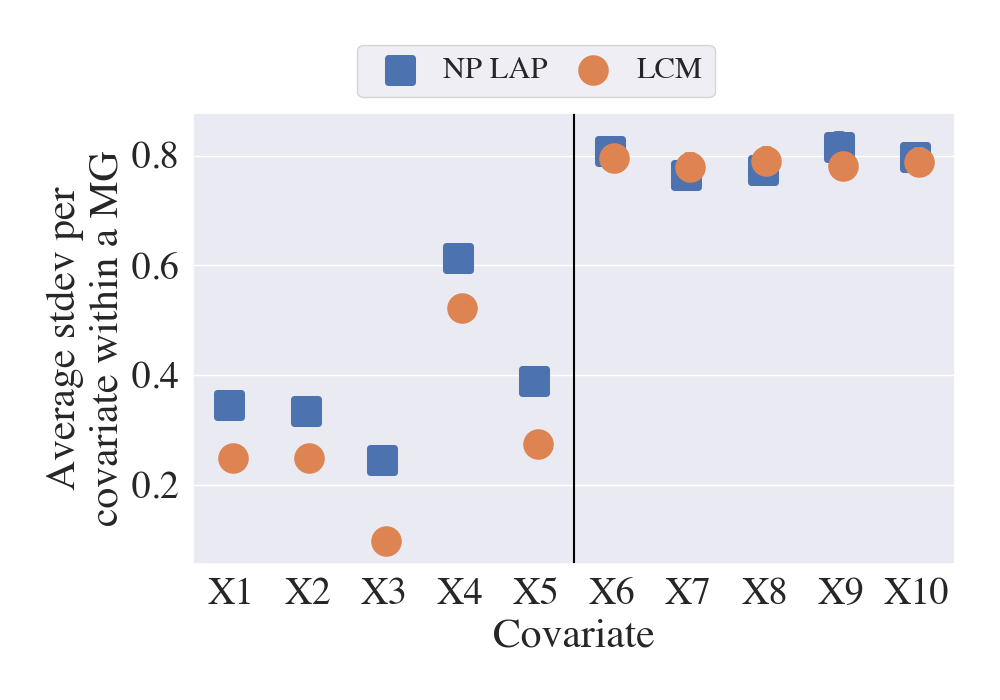}
  \caption{Average standard deviation for each covariate inside the matched groups for non-parametric LAP (NP LAP) and LCM. The smaller the standard deviation, the tighter the match on that covariate. The dataset has 20 covariates, but we only show 10 for ease of presentation. Note that X1-X5 are important and X6-X10 are unimportant.}
  \label{fig:lap_b}
\end{figure}

We compare LCM and LAP using the quadratic data generation process used in Section~\ref{sec:results-scalability} and described in the Supplementary Material. We generate 5000 units and 20 covariates, of which the first 5 are important and the other 15 are irrelevant. Here, we first do 25 nearest neighbors matching with PGM and then perform 5 nearest neighbors matching using the LCM learned distance metric. Figure~\ref{fig:lap_a} shows that for this problem setup, both linear LAP and non-parametric LAP are more accurate than LCM. Further, Figure~\ref{fig:lap_b} shows that the matches created using non-parametric LAP are almost equally as tight as LCM's matches on the 5 important covariates and do not prioritize matching on irrelevant covariates.


\section{Discussion and Conclusion}
Model-to-Match is a fast, scalable, and auditable framework for observational causal inference. Unlike other almost-matching-exactly approaches, Model-to-Match can scale to large datasets and high-dimensional settings and is flexible in regards to how the outcome space is modeled to learn a distance metric. We implemented Model-to-Match using LASSO as our machine learning algorithm of choice and refer to this as LCM. We show many desirable properties of LCM -- including robustness to model misspecification and the ability to handle high-dimensional settings -- and provide details on its consistency for CATE estimation. We provide additional experimental results in the Supplementary Material including further comparisons to non-matching CATE estimation methods and a simulation showing the advantage of LCM over equally weighted matching after feature selection.

\paragraph{Limitations and Future Directions. } Model-to-Match is for i.i.d. data and should be extended to situations with either network interference or time-series effects. Furthermore, Model-to-Match is sensitive to the variable importance metric choice -- leading to confounding bias if the correct support is not recovered. While we introduce our framework for categorical treatments, we are working on extending its application to continuous treatment regimes.

Other variations of Model-to-Match are easily possible. While we show sparse decision trees as a potential substitute to LASSO, any machine learning algorithm can be used. Furthermore, one can use other configurations in the matching and estimation steps of the framework, such as using a $\|\cdot\|_2$ norm instead of $\|\cdot\|_1$, employing a caliper matching method instead of $K$ nearest neighbors, or choosing a different post-matching estimator instead of arithmetic average for potential outcomes. This level of flexibility makes Model-to-Match a framework that can be adapted to a variety of practical problems. In future work, we plan to both study the theory behind different Model-to-Match variations and implement our framework on large, real-world datasets such as electronic health records, genome studies, living standards measurement studies, etc.

\begin{acknowledgements} 
We acknowledge funding from the National Science Foundation and Amazon under grant NSF IIS-2147061, and the National Institute on Drug Abuse under grant DA054994. Quinn Lanners thanks the National Science Foundation Artificial Intelligence for Designing and Understanding Materials - National Research Traineeship (aiM-NRT) at Duke University funded under grant DGE-2022040. Cynthia Rudin, Alexander Volfovsky and Harsh Parikh are supported by NSF grant DMS-2046880. Harsh Parikh is also partially support by Amazon Graduate Fellowship and NSF award IIS-1703431. Alexander Volfovsky is also supported by a National Science Foundation Faculty Early Career Development Award (CAREER: Design and analysis of experiments for complex social processes).
\end{acknowledgements}

\bibliography{lanners_407}

\onecolumn 

\title{Variable Importance Matching for Causal Inference (Supplementary material)}

\maketitle

\section{Proofs for Theorems in Section 5}

\begin{manualtheorem}
{\ref{thm:motivation}}[Closeness in $\X$ implies closeness in $Y$] Consider a $p$-dimensional covariate space where for $t' \in \{0,1\}$, $f^{(t')}(\X_i) = \mathbb{E}[Y_i | \X = \X_i, T = t' ] = \X_i\bm{\beta}^{(t')}$. Construct $\mathcal{M}\in\mathbb{R}^{p\times p}$ where for all $l,r \in \{1,...,p\}$ $\mathcal{M}_{l,l} = |\beta^{(t')}_l|$ and for $l\neq r$ $\mathcal{M}_{l,r} = 0$. Then, $\forall i, j$, we have that $d_{\mathcal{M}}(\X_i, \X_j) \geq \left|f^{(t')}(\X_i) - f^{(t')}(\X_j) \right|$.
\end{manualtheorem}
\textbf{Proof for Theorem~\ref{thm:motivation}}. 
\begin{align*}
    d_{\mathcal{M}}(\X_i, \X_j) = \sum\limits_{l=1}^p \mathcal{M}_{l,l}|X_{i,l} - X_{j,l}| = \sum_{l=1}^p |\beta^{(t')}_l||X_{i,l} - X_{j,l}| &\geq \left|\sum_{l=1}^p \beta^{(t')}_l(X_{i,l} - X_{j,l})\right| \\&= \left|f^{(t')}(\X_i) - f^{(t')}(\X_j) \right|.
\end{align*}
QED

\begin{manualtheorem}{\ref{thm:supp}}[Optimality of $\mathcal{M}$]
    Using the setup of Theorem~\ref{thm:motivation}, let $\supp(\X) = \mathbb{R}^p$.
    Consider an arbitrary diagonal Mahalanobis distance matrix $\widetilde{\mathcal{M}}\in\mathbb{R}^{p\times p}$ where $\|\widetilde{\mathcal{M}}\|_1 = \|\bm{\beta}^{(t')}\|_1$ and $\widetilde{\mathcal{M}}_{l,l} > 0$ when $|\beta^{(t')}_l| > 0$.
    For some $\epsilon \geq 0$ and $\X_1\in\mathbb{R}^p$, define $S_{\widetilde{\mathcal{M}}, \epsilon}(\X_1) := \{\X_2 : \X_2\in\mathbb{R}^p, d_{\widetilde{\mathcal{M}}}(\X_1, \X_2) = \epsilon\}$. Then, 
   \begin{equation*}
        \sup\limits_{\X_2\in S_{\mathcal{M}, \epsilon}(\X_1)}|f^{(t')}(\X_1)- f^{(t')}(\X_2)| 
        \leq
        \sup\limits_{\X_3\in S_{\widetilde{\mathcal{M}}, \epsilon}(\X_1)}|f^{(t')}(\X_1)- f^{(t')}(\X_3)|.
    \end{equation*}
\end{manualtheorem}

In what follows, we recall that a diagonal Mahalanobis distance matrix, $\widetilde{\mathcal{M}}$, is: 
\begin{itemize}
\item  diagonal: for all $l,r \in \{1,...,p\}$, $l\neq r$, $\widetilde{\mathcal{M}}_{l,r} = 0$.
\item non-negative entries: for all $l \in \{1,...,p\}$, $\widetilde{\mathcal{M}}_{l,l} \geq 0$.
\end{itemize}

To prove this result, we first prove the following two lemmas.

\textbf{Lemma 1}\label{lemma-1} (Maximum Absolute Difference in Expected Outcomes under $\mathcal{M}$). Consider a $p$-dimensional covariate space where $\supp(\X) = \mathbb{R}^p$ and for $t' \in \{0,1\}$, $f^{(t')}(\X_i) = \mathbb{E}[Y_i | \X = \X_i, T = t' ] = \X_i\bm{\beta}^{(t')}$. Define $\mathcal{L} := \{l: \left|\beta^{(t')}_l\right| > 0\}$. Construct any diagonal Mahalanobis distance matrix, $\widetilde{\mathcal{M}}$, where $\|\widetilde{\mathcal{M}}\|_1 = \|\bm{\beta}^{(t')}\|_1$ and $\widetilde{\mathcal{M}}_{l,l} > 0$ when $|\beta^{(t')}_l| > 0$. Then, for some $\epsilon\geq 0$ and $\X_1\in\mathbb{R}^p$, let $S_{\widetilde{\mathcal{M}}, \epsilon}(\X_1)$ be as defined in Theorem~\ref{thm:supp}.
We can conclude that 
\begin{equation*}
    \sup\limits_{\X_3\in S_{\widetilde{\mathcal{M}}, \epsilon}(\X_1)}|f^{(t')}(\X_1)- f^{(t')}(\X_3)| = \epsilon \max_{l\in\mathcal{L}}\left\{\frac{|\beta^{(t')}_l|}{\widetilde{\mathcal{M}}_{l,l}}\right\}.
\end{equation*}

\textbf{Proof of Lemma 1}.
\begin{align*}
    \sup\limits_{\X_3\in S_{\widetilde{\mathcal{M}}, \epsilon}(\X_1)}|f^{(t')}(\X_1)- f^{(t')}(\X_3)| &= \sup\limits_{\X_3\in S_{\widetilde{\mathcal{M}},\epsilon}(\X_1)}\left|\sum\limits_{l\in\mathcal{L}} \beta^{(t')}_l(X_{1,l} - X_{3,l})\right|.
\end{align*}
Note that since $\supp(\X) = \mathbb{R}^p$, with probability strictly greater than zero there exists an $\X_1$ and $\X_3$ such that $d_{\widetilde{\mathcal{M}}}(\X_1, \X_3) = \epsilon$ and for all $l\in\mathcal{L}$, $X_{1,l} > X_{3,l}$ when $\beta^{(t')}_l > 0$ and $X_{1,l} < X_{3,l}$ when $\beta^{(t')}_l < 0$. Then,
\begin{align*}
\sup\limits_{\X_3\in S_{\widetilde{\mathcal{M}},\epsilon}(\X_1)}\left|\sum\limits_{l\in\mathcal{L}} \beta^{(t')}_l(X_{1,l} - X_{3,l})\right|
    &= \sup\limits_{\X_3\in S_{\widetilde{\mathcal{M}},\epsilon}(\X_1)}\left\{ \sum\limits_{l\in\mathcal{L}} \left|\beta^{(t')}_l(X_{1,l} - X_{3,l})\right|\right\} \\
    &= \sup\limits_{\X_3\in S_{\widetilde{\mathcal{M}},\epsilon}(\X_1)}\left\{ \sum\limits_{l\in\mathcal{L}} \frac{|\beta^{(t')}_l|}{\widetilde{\mathcal{M}}_{l,l}}\widetilde{\mathcal{M}}_{l,l}\left|X_{1,l} - X_{3,l}\right|\right\}.
\end{align*}
Note that $\left\{\sum\limits_{l\in\mathcal{L}} \frac{|\beta^{(t')}_l|}{\widetilde{\mathcal{M}}_{l,l}}\widetilde{\mathcal{M}}_{l,l}\left|X_{1,l} - X_{3,l}\right| : \X_3\in S_{\widetilde{\mathcal{M}},\epsilon}(\X_1)\right\}$ is maximized at $ \epsilon\max_{l\in\mathcal{L}}\left\{\frac{|\beta^{(t')}_l|}{\widetilde{\mathcal{M}}_{l,l}}\right\}$. It is known that if the maximum value of a set is in the set, the supremum of that set equals the maximum value of that set. Therefore, we conclude that,
\begin{align*}
    \sup\limits_{\X_3\in S_{\widetilde{\mathcal{M}},\epsilon}(\X_1)}\left\{ \sum\limits_{l\in\mathcal{L}} \frac{|\beta^{(t')}_l|}{\widetilde{\mathcal{M}}_{l,l}}\widetilde{\mathcal{M}}_{l,l}\left|X_{1,l} - X_{3,l}\right|\right\}
    &= \epsilon\max_{l\in\mathcal{L}}\left\{\frac{|\beta^{(t')}_l|}{\widetilde{\mathcal{M}}_{l,l}}\right\}.
\end{align*}
QED

\textbf{Lemma 2}\label{lemma-2} Under the same setup as Lemma \hyperref[lemma-1]{1}, $\max_{l\in\mathcal{L}}\left\{\frac{|\beta^{(t')}_l|}{\widetilde{\mathcal{M}}_{l,l}}\right\} \geq 1$.

\textbf{Proof of Lemma 2}.
First note that $\sum\limits_{l\in\mathcal{L}} \widetilde{\mathcal{M}}_{l,l} \leq \sum\limits_{l=1}^p \widetilde{\mathcal{M}}_{l,l} = \sum\limits_{l=1}^p |\beta^{(t')}_l| = \sum\limits_{l\in\mathcal{L}} |\beta^{(t')}_l|$. There are two possible cases. In case one, $\forall l\in\mathcal{L}$, $\widetilde{\mathcal{M}}_{l,l} = \mathcal{M}_{l,l} = |\beta^{(t')}_l|$. Then $\max_{l\in\mathcal{L}} \frac{|\beta^{(t')}_l|}{\widetilde{\mathcal{M}}_{l,l}} = 1$. In case two, there exists $l\in\mathcal{L}$ for which $\widetilde{\mathcal{M}}_{l,l} \neq |\beta^{(t')}_l|$. But then there must exist an $l'\in\mathcal{L}$ for which $\widetilde{\mathcal{M}}_{l',l'} < |\beta^{(t')}_{l'}| \implies \max_{l\in\mathcal{L}} \frac{|\beta^{(t')}_l|}{\widetilde{\mathcal{M}}_{l,l}} > 1$.
QED

\textbf{Proof of Theorem~\ref{thm:supp}}.
First note that $\mathcal{M}$ is a diagonal Mahalanobis distance matrix, $\|\mathcal{M}\|_1 = \|\bm{\beta}^{(t')}\|_1$, and $\mathcal{M}_{l,l} > 0$ when $|\beta^{(t')}_l| > 0$. The proof of the theorem 
then follows directly from Lemma \hyperref[lemma-1]{1} and Lemma \hyperref[lemma-2]{2}.
\begin{equation*}
    \begin{split}
        \sup\limits_{\X_2\in S_{\mathcal{M}, \epsilon}(\X_1)}|f^{(t')}(\X_1)- f^{(t')}(\X_2)|
        &= \epsilon \max_{l\in\mathcal{L}}\left\{\frac{|\beta^{(t')}_l|}{\mathcal{M}_{l,l}}\right\} \\
        &= \epsilon \max_{l\in\mathcal{L}}\left\{\frac{|\beta^{(t')}_l|}{|\beta^{(t')}_l|}\right\} \\
        &= \epsilon \\
        &\leq \epsilon \max_{l\in\mathcal{L}}\left\{\frac{|\beta^{(t')}_l|}{\widetilde{\mathcal{M}}_{l,l}}\right\} \\
        &= \sup\limits_{\X_3\in S_{\widetilde{\mathcal{M}}, \epsilon}(\X_1)}|f^{(t')}(\X_1)- f^{(t')}(\X_3)|.
    \end{split}
\end{equation*}
Where $\epsilon \leq \epsilon \max_{l\in\mathcal{L}}\left\{\frac{|\beta^{(t')}_l|}{\widetilde{\mathcal{M}}_{l,l}}\right\}$ because of Lemma \hyperref[lemma-2]{2}.
QED

\begin{manualtheorem}{\ref{thm:consistency}}[Consistency of LCM]
     For $t'\in\{0,1\}$, let $f^{(t')}(\X_i) = \mathbb{E}[Y_i | \X = \X_i, T = t' ] $. Let $f^{(t')}$ be Lipschitz continuous and,
     \begin{equation*}
        \supp\left(f^{(t')}\right) :=  \left\{j : 
             \textrm{importance of }\X_{\cdot, j}\textrm{ in } f^{(t')}\textrm{ is } > 0\right\}.  
     \end{equation*}
     Denote $d_{\mathcal{M}^*}$ as the distance metric learned by LCM in Section~\ref{sec: method} and let $\Gamma\left(\mathcal{M}^*\right) = \{j : \mathcal{M}^*_{j,j} > 0\}$. LCM is consistent for CATE estimation if $\supp\left(f^{(0)}\right) \bigcup \supp\left(f^{(1)}\right) \subseteq \Gamma\left(\mathcal{M}^*\right)$.
\end{manualtheorem}     
\textbf{Proof of Theorem \ref{thm:consistency}}. 
First, let us introduce the concept of a smooth distance metric (defined in \cite{malts}). 
\begin{definition}[Smooth Distance Metric] \label{def-1}
    $d: \X \times \X \rightarrow \mathbb{R}^+$ is a smooth distance metric if there exists a monotonically increasing bounded function $\delta_d(\cdot)$ with zero intercepts, such that $\forall i, j \in \mathcal{S}$ if $T_i = T_j = t'$ and $d(\X_i, \X_j) \leq a$ then $\left|\mathbb{E}\left[Y_i(t') | \X_i \right] - \mathbb{E}\left[Y_j(t') | \X_j \right]\right|\leq \delta_{d}(a).$
\end{definition}
Theorem 1 in \citep{malts} shows that matching with a smooth distance metric guarantees consistency of CATE estimates. 

Recovering the correct support for the potential outcome functions implies that restricting to only variables in the recovered support, the potential outcomes are independent of the covariates: 
$(Y(1),Y(0)) \perp \X\mid \{\X_{\cdot,j}\}_{j\in \supp( f^{(0)} ) \cup \supp( f^{(1)} )} $. Also, note that if $\{\X_{i,j}\}_{j\in \supp( f^{(0)} ) \cup \supp( f^{(1)} )}$ is close to $\{\X_{k,j}\}_{j\in \supp( f^{(0)} ) \cup \supp( f^{(1)} )}$ then $f^{(0)}(X_i)$ is close to $f^{(0)}(X_k)$ and $f^{(1)}(X_i)$ is close to $f^{(1)}(X_k)$ by the definition of support and the Lipschitz continuity assumption. Thus, if $ \supp( f^{(0)} ) \cup \supp( f^{(1)} )\subseteq \Gamma(\mathcal{M^*})$ then $d_\mathcal{M}^*$ is a smooth distance metric. This guarantees the consistency of our estimates. QED

\paragraph{Consistency of LASSO.} Much work has been done on the consistency of LASSO for feature selection 
\citep{Zhang2016a}.  
The ability for LASSO to recover the correct support even in the case of non-linear targets makes it more robust to model misspecification. LASSO is consistent for support recovery if $f(\X_i, t) = \mathbb{E}[Y_i | \X = \X_i, T = t' ]$ satisfies one of the following conditions:
    \begin{enumerate}[label=(\roman*)]
        \item $f(\X_i, t') = \X_i\bm{\beta^{(t')}}$
        \item $f(\X_i, t') = g\left(\X_i\bm{\beta^{(t')}}\right)$ where $\beta^{(t')}_k \neq 0$ for $k\in\{1,..,r\}$, for some $r\leq p$, and, if $r < p$, $\beta^{(t)}_k = 0$ for $k\in\{r, ..., p\}$, and the following conditions are met:
        \begin{enumerate}
            \item \textbf{Cov}($\X, \X$) is invertible.
            \item The eigenvalues of $\Sigma_{r,r} =$ \textbf{Cov}($\X_{1:r}, \X_{1:r}$) are such that $0 < c_1 \leq \Lambda\left(\Sigma_{r,r}\right) \leq c_2 < \infty$. Where $\Lambda\left(\Sigma_{r,r}\right)$ are the eigenvalues of $\Sigma_{r,r}$.
            \item $E[Y(t')]^4 < \infty$
            \item $g$ is differentiable almost everywhere and for $t\sim\mathcal{N}(0,1)$, $E(|g(t)|) < \infty$ and $E(|g'(t)|) < \infty$.
            \item For all $i$, $E\left[X_i^TX_i\left|g\left(\X_i\bm{\beta^{(t')}}\right)\right|^2\right] < \infty$.
        \end{enumerate}
    \end{enumerate}

\section{Method Implementation for Experiments}
In this section we outline how we implemented each method used in our experiments. To calculate CATE estimates for all samples, we employed the same $\eta$-fold cross-fitting strategy for each method. In particular, we train models to estimate the $\widehat{Y}_i(t') = f^{(t')}(\X_i)$ for $t'\in\{0,1\}$ using $S_{n,tr}$ and perform estimation on $S_{n,est}$. The only method that we did not use cross-fitting for was GenMatch, which does not use the outcome to learn it's distance metric and thus does not require a training set. All references to scikit-learn refer the Python machine learning package from \cite{scikit-learn}.
\begin{itemize}
    \item \textbf{LASSO Coefficient Matching}: We implemented the method described in this paper in Python. We use scikit-learn's \texttt{LassoCV} to learn $d_{\mathcal{M^*}}$ and \texttt{NearestNeighbors} with \texttt{metric='manhattan'} to perform nearest neighbor matching.

    \item \textbf{Linear and Nonparametric Prognostic Score Matching}: We follow the notion of a prognostic score outlined in \cite{Hansen2008}. In particular, we employ a \textit{double} prognostic score matching method were we model both the control and treatment space separately as $\widehat{Y}_i(t') = f^{(t')}(\X_i)$ for $t'\in\{0,1\}$. For linear PGM we use scikit-learn's \texttt{LassoCV} as our prognostic score models and for nonparametric PGM we use \texttt{GradientBoostingRegressor} for our prognostic score models. We then match with replacement on $[f^{(0)}(\X_i), f^{(1)}(\X_i]$ using scikit-learn's \texttt{NearestNeighbors} with \texttt{metric='euclidean'} to perform nearest neighbor matching. We estimated CATEs with the same mean estimator as LCM.

    \item \textbf{MALTS Matching}: We use the method developed in \citet{malts} that was implemented in Python \citep{git_malts}. We use the package's \texttt{mean} CATE estimator with \texttt{smooth\_cate=False}.     

    \item \textbf{MatchIt}: We use MatchIt's implementation of GenMatch \citep{ho2007matching}. We kept the default setting of \texttt{ratio=1}, which set $K=1$ for matching. But we matched with replacement to be in line with LCM and the other matching methods we compared with.
    \item \textbf{Linear and Nonparametric TLearner}: We use the EconML TLearner implementation from \cite{econml}. For Linear TLearner we use scikit-learn's \texttt{LassoCV} for our models and for Nonparametric TLearner we use scikit-learn's \texttt{GradientBoostingRegressor} for our models.

    \item \textbf{AHB}: We use the method developed in \citet{morucci2020adaptive} that was implemented in R \citep{git_ahb}. We use the package's \texttt{AHB\_fast\_match} implementation with the default settings.

    \item \textbf{Bart T-Learner}: We use the dbarts R package from \cite{dbart}. We train  a BART model on $S_{n,tr}$ to model $\widehat{Y}_i(t') = f^{(t')}(\X_i)$ for $t'\in\{0,1\}$. We then estimate CATEs for each $j\in S_{n,est}$ as $f^{(1)}(\X_j) - f^{(0)}(\X_j)$.

    \item \textbf{Linear DoubleML}: We use the \texttt{econml.dml.DML} class in the econml Python package from \cite{econml}. We fit a model on $S_{n,tr}$ setting \texttt{model\_y=WeightedLassoCV}, \texttt{model\_t=LogisticRegressionCV}, and \texttt{model\_final=LassoCV}. We then estimate CATEs for each $j\in S_{n,est}$ using the \texttt{.effect()} method.

    \item \textbf{Causal Forest DoubleML}: We use the \texttt{econml.dml.CausalForestDML} class in the econml Python package from \cite{econml}. We fit a model on $S_{n,tr}$ setting \texttt{model\_y=WeightedLassoCV} and \texttt{model\_t=LogisticRegressionCV}. We then estimate CATEs for each $j\in S_{n,est}$ using the \texttt{.effect()} method.    

    \item \textbf{Causal Forest}: We use the implementation of causal forest from the grf R package from \cite{econml}. We fit a model on $S_{n,tr}$ with the default package settings. We then used the fit model to estimate CATEs for each $j\in S_{n,est}$.

\end{itemize}

\section{Experimental Details for Section~\ref{sec: results} and Section~\ref{sec: extensions}}
In this section, we describe the data generating processes used and provide further details regarding the setup of each experiment conducted in this paper. The source code necessary to reproduce all of the experiments in this paper is located in the GitHub repository: \url{https://github.com/almost-matching-exactly/variable_imp_matching}.

\subsection{Data Generation Processes}\label{sec:dgps}
Here we outline the data generation processes (DGPs) not fully outlined in the main text.

\textbf{Sine and Exponential DGPs}. \textit{Used in Sections~\ref{sec:results-nonlinear} and \ref{sec: metalearner}}. We generate the covariates and treatment assignments for the Sine and Exponential DGPs in a similar manner.
For both, we generate data as follows:
\begin{align*}
    X_{i,1},\dots, X_{i,p} \overset{iid}{\sim} \text{Uniform}(-\alpha, \beta) \\
    \epsilon_{i,y} \overset{iid}{\sim} \mathcal{N}(0, \sigma^2), \epsilon_{i,t} \overset{iid}{\sim} \mathcal{N}(0, 1) \\
    T_i = \mathbbm{1}\Bigg[\text{expit}\Big(X_{i,1} + X_{i,2} + \epsilon_{i,t}\Big) > 0.5\Bigg]  \\
    Y_i = T_i Y_i(1) + (1-T_i) Y_i(0) + \epsilon_{i, y},
\end{align*}
where expit is the logistic sigmoid: \(\text{expit}(x) = \frac{1}{1 + e^{-x}}\).

For \textbf{Sine} we set 
\(\alpha=\beta=\pi\), \(\sigma^2=0.1\) and calculate the potential outcomes as
\begin{equation*}
    Y_i(0) = \sin(X_{i,1}),\; Y_i(1) = \sin(X_{i,1}) - \sin(X_{i,2}).
\end{equation*}
For \textbf{Exponential} we set \(\alpha=\beta=3\), \(\sigma^2=1\) and calculate the potential outcomes as 
\begin{equation*}
    Y_i(0) = 2e^{X_{i,1}} - \sum_{j=2}^3 e^{X_{i,j}},\; Y_i(1) = 2e^{X_{i,1}} - \sum_{j=2}^3 e^{X_{i,j}} + e^{X_{i,4}}.
\end{equation*}

\textbf{Quadratic DGP}. \textit{Used in Sections~\ref{sec:results-scalability} and \ref{sec:lcm-aug-pgm}}. 
This quadratic data generation process is also described in \cite{malts}. This DGP includes both linear and quadratic terms. For each sample, let $\X_{i}$ be a $p$-dimensional vector where the first $k\leq p$ covariates are relevant and $\kappa \leq k$ is the number of covariates relevant to determining the treatment choice. The DGP is outlined below.

\begin{equation*}
    \begin{gathered}
        X_{i,p} \overset{iid}{\sim} \mathcal{N}(1, 1.5), \  \epsilon_{i,y} \epsilon_{i,t} \overset{iid}{\sim} \mathcal{N}(0,1), \
        \ s_1,\dots,s_{|k|} \overset{iid}{\sim} \text{Uniform}\{-1,1\} \\
        \alpha_j|s_j \overset{iid}{\sim} \mathcal{N}(10s_j,9), \ \beta_1,\dots,\beta_{|k|} \overset{iid}{\sim} \mathcal{N}(1,0.25)        
    \end{gathered}
\end{equation*}

\begin{equation*}
    Y_i(0) = \sum_{j\leq k} \alpha_jX_{i,j}    
\end{equation*}
\begin{equation*}
    Y_i(1) = \sum_{j\leq k} \alpha_jX_{i,j} + \sum_{j\leq k}\beta_jX_{i,j} + \sum_{j\leq k} \sum_{j'\leq k} X_{i,j}X_{i,j'}
\end{equation*}
\begin{equation*}
    T_i = \mathbbm{1}\Bigg[\text{expit}\Big(\sum_{j\leq \kappa} X_{i,j} - \kappa +  \epsilon_{i,t}\Big)  > 0.5 \Bigg]
\end{equation*}
\begin{equation*}
    Y_i = T_iY_i(1) + (1-T_i) Y_i(0) + \epsilon_{i,y}
\end{equation*}
Where expit$(x)$ = $\frac{1}{1+e^{-x}}$.

\textbf{Basic Quadratic DGP}. \textit{Used in Section~\ref{sec:feature-imp}}. This DGP is a quadratic DGP centered at zero. We generate each sample as shown.
\begin{equation*}
    X_{i,1},\dots, X_{i,10} \overset{iid}{\sim} \mathcal{N}(0, 2.5), \ \epsilon_{i,y} \overset{iid}{\sim} \mathcal{N}(0, 1), \ T_i \sim\text{Bernoulli}(0.5)
\end{equation*}
\begin{equation*}
    Y_i(0) = X_{i,1}^2, \ Y_i(1) = X_{i,1}^2+10
\end{equation*}
\begin{equation*}
    Y_i = T_i Y_i(1) + (1-T_i)Y_i(0) + \epsilon_{i, y}
\end{equation*}

\subsection{Experimental Details}\label{sec:add-exp-details}
In Table~\ref{tab:exp-details} we provide details on the experiments shown in this paper. We include additional notes for selected experiments below:

\begin{itemize}
    \item Section~\ref{sec:results-auditability}: Accuracy and
Auditability: We included the school id as a categorical covariate in our dataset. After preprocessing the categorical covariates, we had 6 continuous covariates and 98 binary covariates that we used as input to each model. We used only two splits due to the small occurrence rate of many of the categorical values. We repeated the cross-fitting process 50 times to smooth out treatment effect estimates for each method. All of the results in this section are for the combined 50 iterations.
    \item Section~\ref{sec:results-scalability}: Scalability: The matchit package only performs k:1 matching, so we kept K=1 for GenMatch (which is the default value). Reported runtimes were measured on a Slurm cluster with VMware, where each VM was an Intel(R) Xeon(R) CPU E5-2699 v4 @ 2.20GHz. For measuring runtime, we ran each method 20 times on each dataset size. We report the average runtime for each method on each dataset. The variability across the 20 runs was negligible so we ommitted bars showing the standard deviation from the final plot. Each individual runtime measurement was ran on a separate Slurm job that was allocated a single core with 16GB RAM.
    \item Section~\ref{sec:lcm-aug-pgm}: LCM-Augmented-PGM: For ease of implementation, we did not perform cross-fitting for this experiment. Rather, we just used half of the samples (2500) for training and the other half of the samples (2500) for estimation.
\end{itemize}

\begin{table}[]
\caption{Details of Experiments in Sections \ref{sec: results} and \ref{sec: extensions}. The \textit{Additional Information} column indicates if further details for that experiment are included in Section~\ref{sec:add-exp-details}.}
\centering
\begin{tabular}{|l|l|l|l|l|l|l|}
\hline
\textbf{Section}                                                                                             & \textbf{Dataset}                                                             & \textbf{\# Samples} & \textbf{\# Covariates} & \textbf{K}                                                                   & \textbf{$\eta$} & \textbf{Additional Notes} \\ \hline
\begin{tabular}[c]{@{}l@{}}\ref{sec:results-auditability}: Accuracy and\\ Auditability\end{tabular}  & \begin{tabular}[c]{@{}l@{}}ACIC 2018 Learning\\ Mindset Dataset\end{tabular} & 10,000              & 10                     & 10                                                                           & 2            & Y                         \\ \hline
\multirow{2}{*}{\begin{tabular}[c]{@{}l@{}}\ref{sec:results-nonlinear}: Nonlinear \\ Outcome\end{tabular}} & Sine                                                                         & 5000                & 100                    & 10                                                                           & 10           &                           \\ \cline{2-7} 
                                                                                                             & Exponential                                                                  & 5000                & 100                    & 10                                                                           & 10           &                           \\ \hline
\ref{sec:results-scalability}: Scalability                                                                   & Linear + Quadratic                                                           & Varies              & Varies                 & \begin{tabular}[c]{@{}l@{}}10 (1 for \\ GenMatch\\ - see notes)\end{tabular} & 2            & Y                         \\ \hline
\begin{tabular}[c]{@{}l@{}}\ref{sec: metalearner}: Metalearner \\ LCM\end{tabular}                    & Sine                                                                         & 500                 & 10                     & 10                                                                           & 5            &                           \\ \hline
\begin{tabular}[c]{@{}l@{}}\ref{sec:feature-imp}: Feature\\ Importance Matching\end{tabular}         & Simple Quadratic                                                             & 500                 & 10                     & 10                                                                           & 5            &                           \\ \hline
\begin{tabular}[c]{@{}l@{}}\ref{sec:lcm-aug-pgm}: LCM-\\ Augmented-PGM\end{tabular}                  &   Linear + Quadratic                                                                           &        5000             &             20           &        \begin{tabular}[c]{@{}l@{}}25 using PGM\\ followed by\\ 5 using LCM\end{tabular}                                                                     &   N/A       &  Y \\ \hline
\end{tabular}
\label{tab:exp-details}
\end{table}

\section{Additional Experimental Results}
In this section, we include additional experimental results using LCM. We first discuss further findings from experiments in Section~\ref{sec: results} and Section~\ref{sec: extensions}. We then show results of additional experiments comparing LCM to non-matching methods and matching methods with equal weights after feature selection.

\begin{figure}
\centering
\includegraphics[width=0.6\linewidth]{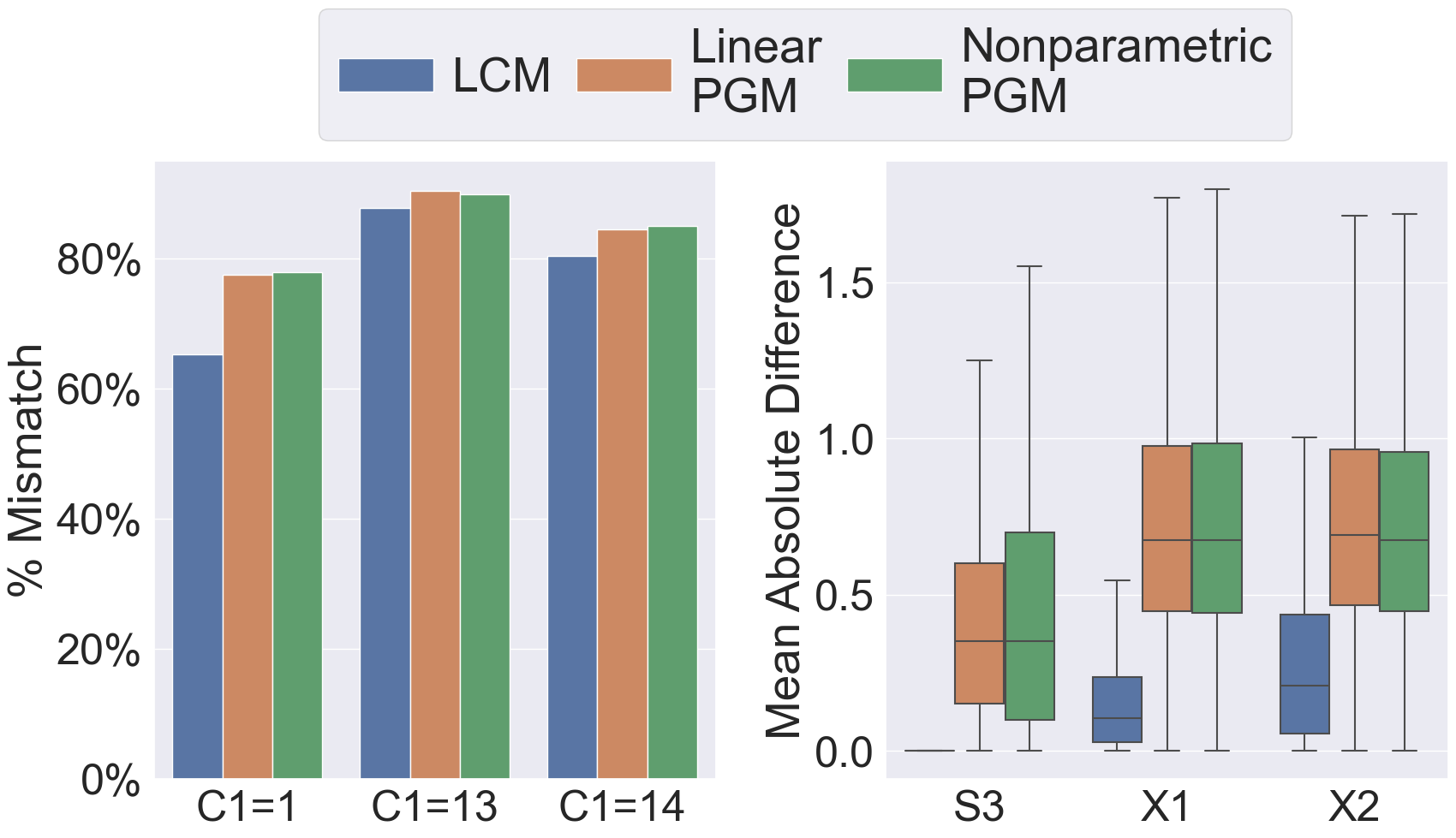}
  \caption{Closeness in important covariates for matched groups produced by LCM, linear PGM, and nonparametric (NP) PGM. Smaller values imply better and tighter matches.}
  \label{fig:schools-mg-full}
\end{figure}

\textbf{Section~\ref{sec:results-auditability}: Accuracy and Auditability}. Figure~\ref{fig:schools-mg-full} in this document is an expanded plot of Figure~\ref{fig:schools-mg-diff}(a) in the main text. The supplementary material's Figure~\ref{fig:schools-mg-full} includes S3, X1, and all other effect modifiers X2, C1=1, C1=13, and C1=14. As mentioned in the caption of Figure~\ref{fig:schools-mg-diff}(a) in the main text, $\text{S3}$ indicates the self-reported prior achievements of students and $\text{X1}$ indicates school-level average mindset score of the students. X2 is a school-level continuous covariate that measures the school's achievement level and C1 is a categorical covariate for race/ethnicity. We measure closeness in continuous covariates using the same mean absolute difference metric used in Figure~\ref{fig:schools-mg-diff}(a) in the main text. Whereas, we measure closeness in categorical covariates as the percent of samples in a match group that do not have the same label as the query unit (\% Mismatch). LCM matches much more tightly on all of the continuous covariates. For categorical covariates, while LCM matches tighter than PGM methods, it struggles compared to continuous covariates. We theorize this is due to the low occurrence rate of these features. In particular, C1=1 in 9.5\%, C1=13 in 1.8\% and C1=14 in 6.2\% of samples. Therefore, it is difficult to find matches that have the same C1 value and are also similar in all of the other important covariates. LCM sometimes prioritizes matching almost-exactly on other covariates at the expense of these rare categorical covariates.

\begin{figure}
\centering
\includegraphics[width=0.6\linewidth]{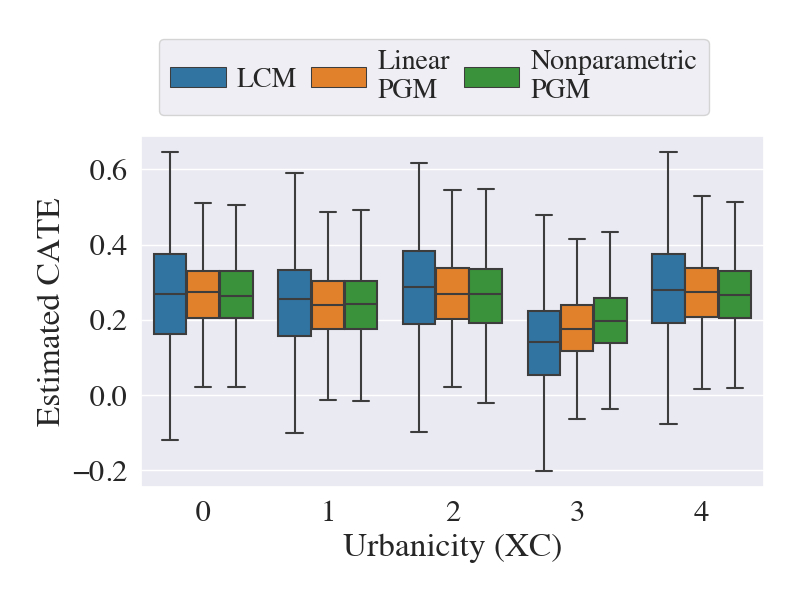}
  \caption{Marginal CATE estimates produced by LCM, Linear PGM, and Nonparametric PGM for the categorical school-level covariate of urbanicty (XC).}
  \label{fig:schools-xc}
\end{figure}

\cite{Carvalho2019} also states that although XC (Urbanicity) is not an effect modifier it is strongly related to X1 (student's fixed mindsets - summarized at the school level) and X2 (school achievement level) which are true effect modifiers. Because of this, seven of the eight methods that are summarized in \cite{Carvalho2019} identified XC as an effect modifier. \cite{Carvalho2019} further shows that, in this dataset, marginally the true cates for XC=3 are much lower than other values of XC. We show in Figure~\ref{fig:schools-xc} that LCM also identifies this trend in XC.

For Section~\ref{sec:results-auditability}, we did not compare to other almost-matching-exactly methods (i.e. MALTS, AHB, GenMatch) due to the large size of the dataset. The ACIC 2018 Learning Mindset Dataset has 50,000 samples and >100 covariates after encoding the categorical features. Results from Section~\ref{sec:results-scalability} highlight how intractable it would be to run other AME methods on a dataset of this size.

\begin{figure}
\centering
\includegraphics[width=0.6\linewidth]{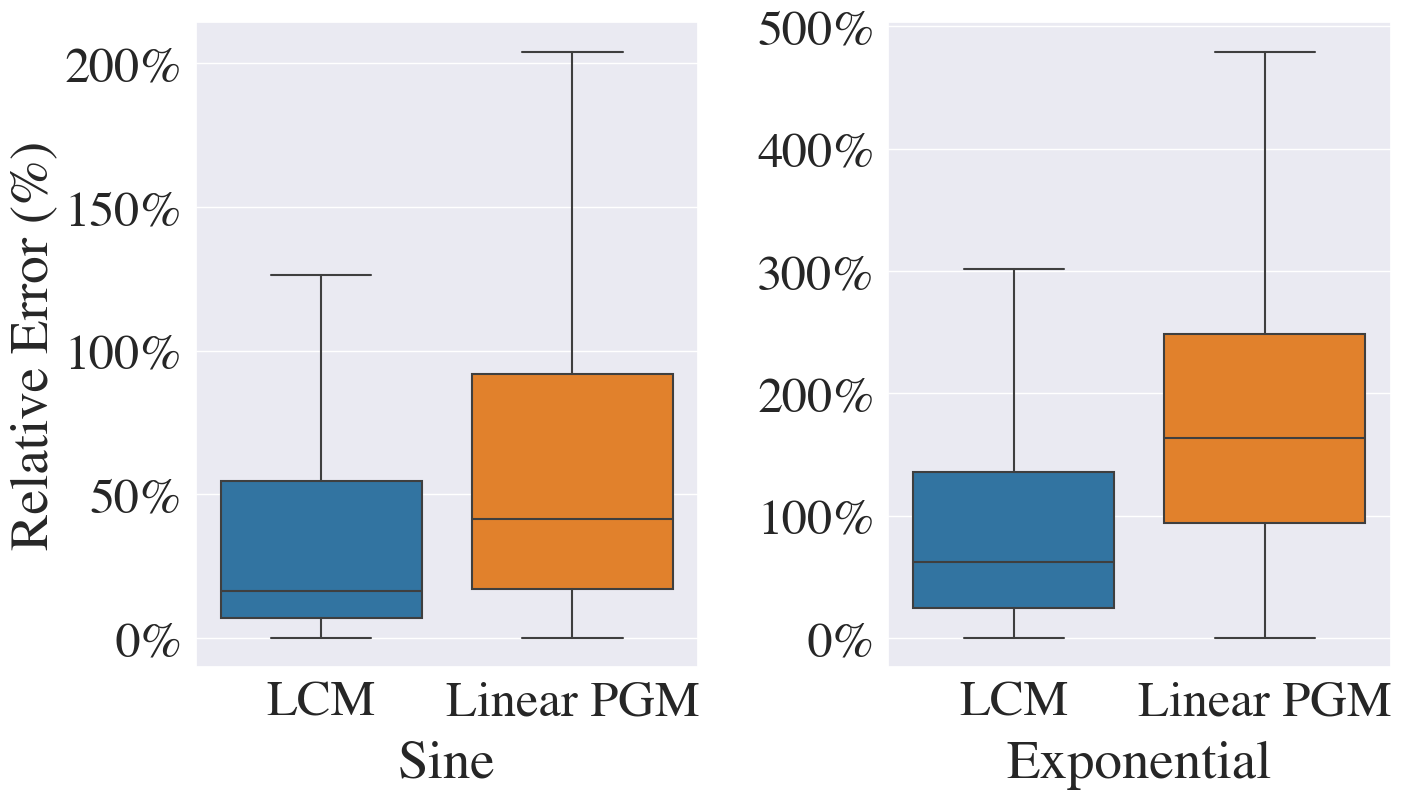}
    \caption{Comparing LCM’s and Linear PGM’s performances for high-dimensional nonlinear synthetically generated datasets \textbf{Sine} and \textbf{Exponential}.}
  \label{fig:nonlinear-highdim}
\end{figure}

\textbf{Section~\ref{sec:results-nonlinear}: Nonlinear Outcomes}. Figure~\ref{fig:nonlinear-highdim} shows CATE estimation accuracy for the same experiment in Section~\ref{sec:results-nonlinear} with the number of covariates increased to 500 for both the \textbf{Sine} and \textbf{Exponential} datasets. Given that we used 10 splits for this experiment, the training set in each fold had 500 samples. Note that LCM's accuracy does not suffer in this extremely high-dimensional setting where the number of samples equals the number of covariates. These results further highlight the ability of LCM to scale to very high-dimensional data even in the case of nonlinear outcome functions.

\begin{figure}
\centering
\includegraphics[width=0.5\linewidth]{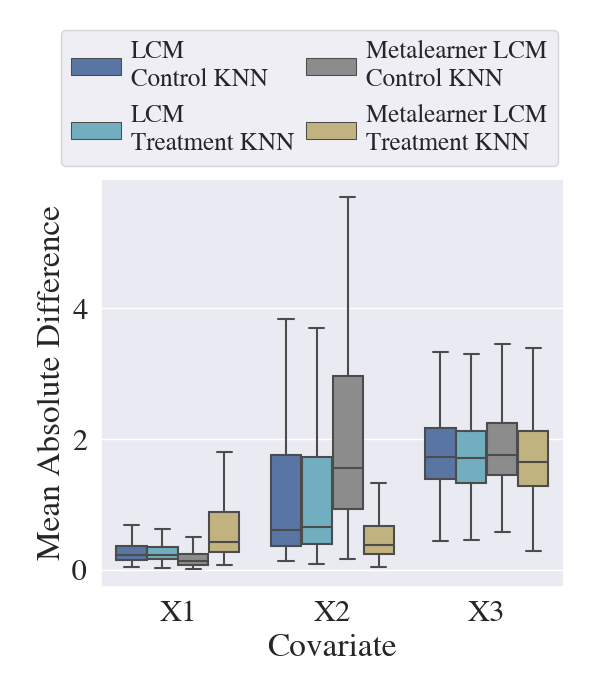}
  \caption{Measure of how tightly the KNN groups are for LCM versus Metalearner LCM under different treatment regimes.}
  \label{fig:metalearner-mg-diff}
\end{figure}

\textbf{Section~\ref{sec: metalearner}: Metalearner LCM}. For the Metalearner LCM, here we show the effect of learning unique distance metrics for calculating control vs treated KNNs. We measure the distance between query unit's covariate values and the values of the ten nearest neighbors' of each treatment type. In particular, we calculate the mean absolute difference between a query unit's value and the values of its ten nearest neighbors. As explained in Section~\ref{sec: metalearner}, X1 is a relevant covariate to the outcome under both treatment regimes, whereas X2 is only relevant to the outcome under treatment. X3 is unimportant in both setting and shown as a reference point. Figure~\ref{fig:metalearner-mg-diff} shows that while LCM's nearest neighbors are equally close on X0 and X1 in both treatment spaces, Metalearner LCM considers X2 as unimportant when calculating KNNs who are in the control group. This highlights how Metalearner LCM is able to adapt to outcome spaces that are different under different treatment regimes.

\textbf{LCM vs Machine Learning Methods}. Previous almost-matching-exactly literature has established that AME methods perform as well as (and often better than) machine learning methods like BART, causal forest, and double machine learning for estimating CATEs \citep{malts, morucci2020adaptive, wang2017flame}. For this reason, this paper focuses on comparing LCM to matching methods and particularly other AME methods. However, here we include an experiment comparing the CATE estimation accuracy of LCM to various machine learning methods on a high-dimensional non-linear dataset. 

\begin{figure}
\centering
\includegraphics[width=0.6\linewidth]{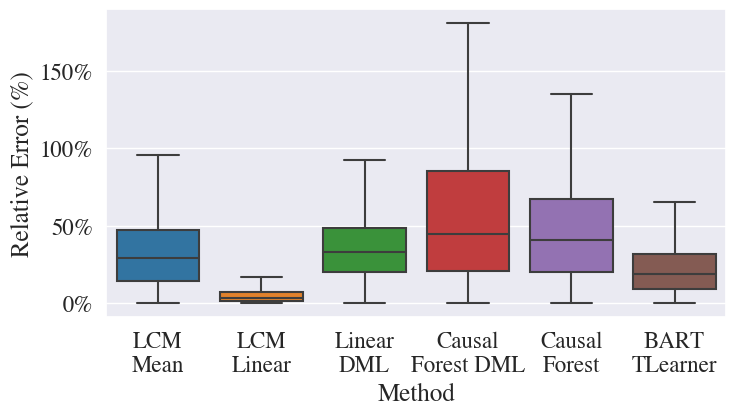}
    \caption{Estimated CATE absolute error relative to the true ATE for LCM Mean, LCM Linear, and state-of-the-art machine learning methods. DML stands for double machine learning.}
  \label{fig:lcm-vs-ml}
\end{figure}

We use the Quadratic DGP with 25 relevant covariates, 2 of which are relevant to the treatment choice, and 125 irrelevant covariates. We generate 2500 samples and set $\eta=5$. We run LCM with two configurations. \textit{LCM Mean} is run with $K=10$ and uses a mean estimator inside the match groups. \textit{LCM Linear} is run with $K=40$ and uses linear regression as the estimator inside the match groups. We compare to state-of-the-art machine learning methods double machine learning (DML), causal forest, and BART TLearner. Figure~\ref{fig:lcm-vs-ml} shows that LCM Mean performs on par with the machine learning methods on this dataset, further highlighting the accuracy our method. LCM Linear improves upon LCM Mean, showing that we can achieve better accuracy with more sophisticated estimators if we are willing to increase the size of the match groups.

\textbf{LCM vs Feature Selection}. Here we show CATE estimation accuracy of LCM compared to matching equally on the covariates after feature selection. To compare with LCM, we estimate CATEs using feature selection by simply following the same steps as LCM but replacing the $\mathcal{M}^*$ with an $\mathcal{M}\in\mathbb{R}^{p\times p}$ such that $\mathcal{M}_{l,l} = 1$ when $\mathcal{M}^*_{l,l} > 0$ and $\mathcal{M}_{l,l} = 0$ when $\mathcal{M}^*_{l,l} = 0$. We refer to this method as \textit{LASSO FS}. We also compare to an \textit{Oracle} feature selector in which we assume that we know which covariates are important and match equally only on the important covariates.

\begin{figure}
\centering
\includegraphics[width=0.9\linewidth]{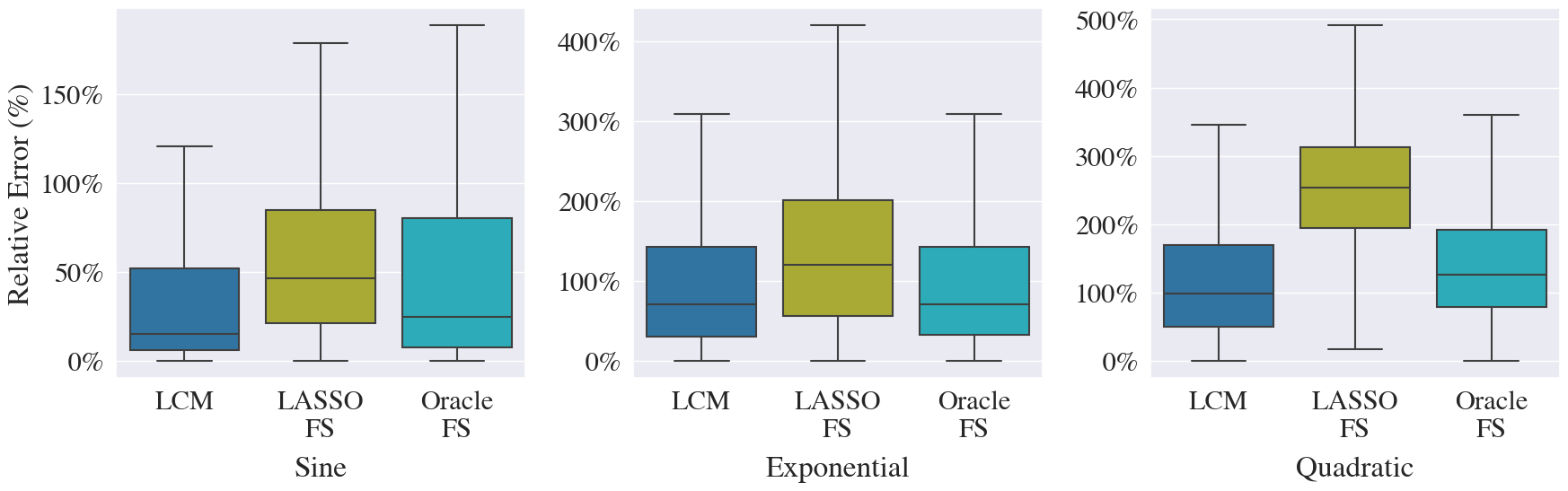}
  \caption{Estimated CATE absolute error relative to the true ATE for LCM and matching equally on covariates after LASSO and Oracle feature selection.}
  \label{fig:lcm-vs-fs}
\end{figure}

We run our analysis on three of the data generation processes used earlier in this paper. Namely, we run on the \textbf{Sine}, \textbf{Exponential}, and \textbf{Quadratic} DGPs described in Section~\ref{sec:dgps}. We generate 5000 samples and 100 covariates for each DGP and have two important covariates for \textbf{Sine}, four important covariates for \textbf{Exponential}, and five important covariates for \textbf{Quadratic}. All tests set $\eta=5$ and $K=10$. Figure~\ref{fig:lcm-vs-fs} shows that LCM outperforms LASSO feature selection and performs on par with an Oracle feature selector. This highlights how using the relative weights of feature importance values in a distance metric, and thus matching tighter on covariates that more heavily contribute to the outcome, ultimately leads to more accurate CATE estimates.

\end{document}